\documentstyle[12pt]{article}
\title{$\Delta(1232)$ and the Polarizabilities of the Nucleon$^*$}
\author{Thomas R. Hemmert$^a$, Barry R. Holstein,$^b$
and Joachim Kambor$^c$\\
{\small $^a$ Theory Division, TRIUMF, 4004 Wesbrook Mall}\\
{\small Vancouver, BC V6T 2A3, CANADA}\\
{\small $^b$ Department of Physics and Astronomy, 
University of Massachusetts}\\
{\small Amherst, MA  01003, USA}\\
{\small $^c$ Theoretische Physik, Universit\"at Z\"urich}\\
{\small CH-8057 Z\"urich, Switzerland}}
\begin{document}
\maketitle
\vfill
\hspace{10cm} ZU-TH-34/96 

\hspace{10cm} TRI-PP-96-68
\begin{titlepage}
\begin{abstract}
Previous calculations of the polarizabilities of the nucleon within the
framework of heavy baryon chiral perturbation theory have included the
contribution of the $\Delta(1232)$ only in its effects on various
contact terms or have been performed in chiral SU(3) where 
systematic errors are difficult to control.  Herein we perform a corresponding
calculation in chiral SU(2)
wherein $\Delta$(1232) is treated as an explicit degree of freedom and
the expansion is taken to third order in soft momenta, the pion mass
and the quantity $M_\Delta-M_N$, collectively denoted $\epsilon$. We present 
the results of a systematic ${\cal O}(\epsilon^3)$ calculation of forward 
Compton scattering off the nucleon, extract the electric
polarizability $\bar{\alpha}_E$, the magnetic polarizability $\bar{\beta}_M$
and the spin polarizability $\gamma$ and compare with available information
from experiments and from previous calculations. Concluding with
a critical discussion of our results, we point out the necessity of a future 
${\cal O}(\epsilon^4)$ calculation.
\end{abstract}
\vfill
$^*$ Research supported in part by the National Science and Engineering 
Research Council of Canada, by the U.S. National Science Foundation and
by Schweizerischer Nationalfonds.
\end{titlepage}
\section{Introduction}
Understanding of the implications of QCD within the regime 
of low energy physics has during
the past decade become accessible via the technique of chiral
perturbation theory (ChPT) \cite{GL84}.  Initial applications 
were in the arena of Goldstone boson interactions \cite{DGH} together with a 
few calculations in 
the baryon sector \cite{BSW85,GSS88}, using relativistic baryon ChPT. 
In recent years use of so-called heavy baryon
methods \cite{G90,JM91a} has generated much interest in calculations 
involving baryons and a great deal of work has 
been done studying strong, weak, and electro-magnetic physics in the near
threshold region \cite{BKM95}. In this work we will focus on 
SU(2) heavy baryon chiral perturbation theory (HBChPT), which has become the
most fully developed sector within baryon ChPT \cite{BKM95,BKKM92,Ecker96}. 
Thus-far, consistent extension to higher energies in SU(2) HBChPT has been 
limited by treatment of the important $\Delta(1232)$ resonance only in
terms of its contribution to the various counterterms which arise in
such calculations. The technique by which to address this deficiency
was developed some time ago \cite{JM91} by including the $\Delta(1232)$ 
as an explicit degree of freedom in a chiral perturbative
scheme\footnote{Using the formalism of \cite{JM91}, Butler and Savage
\cite{BS92} have given an estimate of the contribution of the spin 3/2
resonances to the electric and magnetic polarizabilities of the nucleon.  
However, this calculation was performed in SU(3) and made a number of
 approximations,
so that a direct comparison with systematic SU(2) work is not possible.}.  
Recently, a reformulation of this formalism
which allows for a systematic and explicit calculation of higher order
terms in an expansion of soft momenta, the pion mass and the mass
difference $m_\Delta-m_N$ has been given,\cite{HHK96,HHK96a} and in this note 
we apply this technique to the problem of
forward nucleon Compton scattering and the polarizabilities of the nucleon.
Nucleon Compton scattering is an
area of research which has recently received a great deal of
attention, both experimentally and theoretically and in the next
section we review the status of such work.  In section 3 we give
a brief introduction to the formalism necessary to include the
$\Delta(1232)$ in chiral calculations, and in section 4 apply this to
evaluate delta contributions to $N-\gamma$ scattering,
examine its influence on the polarizabilities and give a critical discussion
of our ${\cal O}(\epsilon^3)$ results.  Finally, in a concluding section 5 we
summarize our findings.

\section{Compton Review}

To lowest order the spin-averaged amplitude for Compton scattering on
the nucleon is given by the Thomson amplitude
\begin{equation}
{\rm Amp}=-{Q^2\over M}\hat{\epsilon}\cdot\hat{\epsilon}' \label{eq:1}
\end{equation}
where $Q,M$ represent the nucleon charge, mass and
$\hat{\epsilon},\hat{\epsilon}'$ and
$k_\mu=(\omega,\vec{k}),{k'}_\mu=(\omega',\vec{k}')$ 
specify the polarization
vectors and four-momenta of the initial,final photons respectively.  In
next order are generated contributions arising from electric and
magnetic polarizabilities---$\bar{\alpha}_E$ and
$\bar{\beta}_M$---which
measure the response of the nucleon to the application of quasi-static
electric and magnetic fields
\begin{equation}
{\rm Amp}=\hat{\epsilon}\cdot\hat{\epsilon}'\left(-{Q^2\over
M}+\omega\omega' \; 4\pi\bar{\alpha}_E \right)
+\hat{\epsilon}\times\vec{k}\cdot\hat{\epsilon}'
\times\vec{k}'\;4\pi\bar{\beta}_M+\; {\cal O}(\omega^4) \; .
\end{equation}
The associated differential scattering cross section on the proton is given by 
\begin{eqnarray}
{d\sigma\over d\Omega}&=&\left({e^2\over 4\pi M}\right)^2
\left({\omega'\over \omega}\right)^2\left[{1\over 2}
(1+\cos^2\theta)\right.\nonumber\\
&-&\left.{4\pi M\omega\omega'\over e^2}\left({1\over 2}
(\bar{\alpha}_E+\bar{\beta}_M)(1+\cos\theta)^2+{1\over
2}(\bar{\alpha}_E-\bar{\beta}_M)(1-\cos\theta)^2\right)+\ldots\right]
\nonumber\\
\quad
\end{eqnarray}
Thus by measurement of the differential Compton scattering cross
section one can extract the electric and magnetic polarizabilities,
provided 
\begin{itemize}
\item[i)] the energy is large enough that these terms are
significant with respect to the Thomson contribution but
\item[ii)]  not so large
that higher order effects dominate.
\end{itemize}
This has been
accomplished for the proton in the energy regime 50 MeV$ <\omega<$100
MeV, yielding\cite{Nat96}
\begin{equation}
\bar{\alpha}_E^p=(11.6\pm 0.6\pm 0.6)\times
10^{-4}{\rm fm}^3;\qquad\bar{\beta}_M^p=(2.6\mp 0.6\mp 0.6)\times
10^{-4}{\rm fm}^3\label{eq:yy}
\end{equation}
(Note: In practice one generally uses the results of
unitarity and the validity of the forward scattering dispersion
relation which yields the Baldin sum rule
\begin{equation}
\bar{\alpha}_E^p+\bar{\beta}_M^p={1\over 2\pi^2}\int_0^\infty
{d\omega\over \omega^2}\sigma_{\rm tot}^p(\omega)=(14.2\pm 0.3)\times
10^{-4}{\rm fm}^3
\end{equation}
as a constraint, since the uncertainty associated with the integral
over the photo-absorption cross section $\sigma_{\rm tot}(\omega)$ is
smaller than that associated with the polarizability measurements.) 

Since the neutron has no charge, such a Thomson-polarizability
interference experiment is not possible, so alternative methods must
be used.  The most precise measurement involves a recent n-Pb
scattering experiment, wherein the dipole moment induced in the moving
neutron due to the nuclear charge acts back on the Pb nucleus.  This
second order process and the resulting $1/r^4$ interaction
proportional to the electric polarizability can be detected in a
transmission experiment via the characteristic linear dependence on
the beam momentum $k$.  This experiment was recently performed at ORNL
and yielded results\cite{Schm91}
\begin{equation}
\bar{\alpha}_E^n=(12.6\pm 1.5\pm 2.0)\times
10^{-4}{\rm fm}^3;\qquad\bar{\beta}_M^n=(3.2\mp 1.5\mp 2.0)\times
10^{-4}{\rm fm}^3\label{eq:zz}
\end{equation}
quite similar to those of the proton. (In this experiment only the electric
polarizability is measured directly. However, using the unitarity sum
rule result
\begin{equation}
\bar{\alpha}_E^n+\bar{\beta}_M^n={1\over
2\pi^2}\int_0^\infty{d\omega\over \omega^2}\sigma_{\rm tot}^n(\omega)=(15.8\pm
0.5)\times 10^{-4}{\rm fm}^3
\end{equation}
the magnetic polarizability can be extracted.)

On the theoretical side, constituent quark-based approaches which rely
on the sum rule\cite{Fr75}
\begin{equation}
\bar{\alpha}_E={1\over 3M}<\sum_{i=1}^3e_i(\vec{r}_i-\vec{R}_{cm})^2>
+2\sum_{n\neq 0}{|<n|\sum_{i=1}^3e_i(\vec{r}_i-\vec{R}_{cm})_z|0>|^2
\over E_n-E_0} \label{eq:aa}
\end{equation}
are bound to fail since the sum over intermediate states component of
Eq.(\ref{eq:aa}) is in almost any reasonable model identical for both
neutron or proton, leading to a prediction
\begin{equation}
\bar{\alpha}_E^p-\bar{\alpha}_E^n\simeq{\alpha\over
3M}(<r_p^2>-<r_n^2>)=4.2\times 10^{-4}{\rm fm}^3
\end{equation}
in conflict with the experimental result that 
$\bar{\alpha}_E^n>\bar{\alpha}_E^p$.\cite{H9}  The solution to the problem lies
in a proper treatment of the pion degrees of freedom of the nucleon,
which is suggested by the feature that the leading pion loop
contributions to neutron and proton are identical, in agreement with
experiment.  The problem can best be addressed using the technique of
heavy baryon chiral perturbation theory \cite{BKM95}, within which at ${\cal
O}(p^3)$ one finds a result \cite{BKKM92}
\begin{equation}
\bar{\alpha}_E^p=\bar{\alpha}_E^n=10\bar{\beta}_M^p=10\bar{\beta}_M^n=
{5\alpha g_A^2\over 96\pi F_\pi^2m_\pi}=12.2\times 10^{-4}{\rm fm}^3
\end{equation}
This ${\cal O}(p^3)$ calculation represents the leading result for 
$\bar{\alpha}_E$ and $\bar{\beta}_M$ in ChPT, but gets the
qualitative features of the polarizabilities right and even agreement
with experiment ! The results diverge as $1/m_\pi$ in the chiral limit, 
giving support to the idea that at these low energies the photon interacts
primarily  with the long-range pion cloud of the nucleon. In order
to understand the experimental finding that $\bar{\alpha}_E^n>\bar{\alpha}_E^p$
the leading ChPT result is not sufficient. One must include higher
order terms in order to find isospin dependent effects and to judge the
convergence behaviour of the series\footnote{Higher order corrections to the
${\cal O}(p^3)$ HBChPT results for the polarizabilities have been discussed
by L'vov \cite{Lvov}.} .
A calculation at ${\cal O}(p^4)$ has been performed by Bernard, Kaiser,
Schmidt and Mei\ss ner \cite{BKSM93}.  At this order four counterterms are
required, which were
estimated by BKSM by treating higher resonances---including the delta 
resonance---as very heavy
with respect to the nucleon.  The results of this process are
\begin{eqnarray}
\bar{\alpha}_E^p&=&(10.5\pm2.0)\times
10^{-4}{\rm fm}^3;\quad\bar{\beta}_M^p=(3.5\pm 3.6)\times
10^{-4}{\rm fm}^3\nonumber\\
\bar{\alpha}_E^n&=&(13.4\pm 1.5)\times 10^{-4}{\rm fm}^3;\quad
\bar{\beta}_M^n=(7.8\pm 3.6)\times 10^{-4}{\rm fm}^3
\end{eqnarray}
where the uncertainty is associated with the counterterm contribution
from the $\Delta$ and from $K,\eta$ loop effects. A very interesting aspect
of this ${\cal O}(p^4)$ calculation lies in the fact that it identifies a 
mechanism to counter the large positive contribution on $\bar{\beta}_M$ due to
$\Delta$(1232) resonance exchange, which is a well-known problem in 
calculations of the magnetic polarizability via effective lagrangians 
\cite{PS95}. BKSM found that at ${\cal O}(p^4)$ (negative) $N\pi$-loop 
contributions can essentially balance out the (positive) Delta effects hidden 
in the counterterms ! Nevertheless, we observe that the uncertainties of their
${\cal O}(p^4)$ calculation are quite dramatic so that
real understanding of the ChPT predictions for the
polarizabilities will require more work. In particular, the large uncertainty 
in $\bar{\beta}_M$ is mainly related to poorly known couplings involving 
$\Delta$(1232), which was used to determine some of the counterterms via
``resonance saturation''. If one retains $\Delta$(1232) as an explicit degree
of freedom in the chiral calculation, as done in our 
${\cal O}(\epsilon^3)$ calculation of section 4, one can in principle 
determine all delta couplings of interest from (other) experiments in a
systematic fashion. Furthermore, one does not limit oneself to the narrow 
scope of ``resonance saturation'' for the spin 3/2 contributions. We refer 
to section 4 for further discussion of this point. 

Finally, with respect to the magnetic polarizability, we note that the
simple quark model {\it can} provide a basic understanding of experiment.  
The prediction\cite{Fr75}
\begin{eqnarray}
\bar{\beta}_M&=&-{1\over 2M}(\sum_ie_i(\vec{r}_i-\vec{R}_{cm})^2>
-{1\over 6}<\sum_ie_i^2(\vec{r}_i-\vec{R}_{cm})^2/m_i>\nonumber\\
&+&2\sum_{n\neq 0}{|<n|\sum_i{e_i\over 2m_i}\sigma_{iz}|0>|^2\over E_n-E_0}
\end{eqnarray}
involves a substantial diamagnetic recoil contribution
\begin{equation}
\bar{\beta}_M^{\rm dia}=\left\{\begin{array}{ll}
-10.2\times 10^{-4}{\rm fm}^3 & p\\
-8.5\times 10^{-4}{\rm fm}^3 & n
\end{array}\right.
\end{equation}
which when added to the large paramagnetic contribution due to the 
$\Delta(1232)$\cite{M9}
\begin{equation}
\bar{\beta}_{M}^\Delta=\left\{\begin{array}{ll}
+12\times 10^{-4}{\rm fm}^3 &p \\
+12\times 10^{-4}{\rm fm}^3 &n 
\end{array}\right.
\end{equation}
gives results in basic agreement with the experimental findings.  
Hence, it is clear that proper inclusion of the $\Delta$ degrees of
freedom is essential. 
\vspace{5mm}

When spin-dependence is included the situation becomes
somewhat more complex.  In order to simplify the present exploratory 
analysis,
we restrict our attention to forward scattering, in which case the
amplitude can be written as
\begin{equation}
{\rm Amp}=f_1(\omega) \; \hat{\epsilon} \cdot \hat{\epsilon}^\prime
+i\vec{\sigma}\cdot\left(\hat{\epsilon}^\prime\times \hat{\epsilon} \right)
\;\omega \; f_2(\omega) \label{eq:f1f2}
\end{equation}
where $f_1(\omega),f_2(\omega)$ are both even functions under
crossing---$\omega\rightarrow -\omega$---and are therefore functions
only of $\omega^2$.  In terms of our previous notation we have
\begin{equation}
f_1(\omega)=-{e^2\over 4\pi M}+(\bar{\alpha}_E+\bar{\beta}_M)\omega^2+\ldots
\end{equation}
while in the case of $f_2(\omega)$ we can make a similar expansion\cite{GMGL}
\begin{equation}
f_2(\omega)=-{e^2\kappa^2\over 2M^2}+\gamma\omega^2+\ldots \label{eq:gamma}
\end{equation}
where $\kappa$ is the anomalous magnetic moment of the target and the
new structure $\gamma$ is the ``spin-polarizability.''  As in the case
of $f_1(\omega)$ the form of the leading term in the expansion is
dictated by rigorous low energy theorems, while the $\omega^2$
correction represents a probe of hadronic structure.  One difference
between the spin-dependent and spin-averaged amplitudes, however, is
the asymptotic behavior as $\omega\rightarrow \infty$.  The better
behavior of $f_2(\omega)$ suggested by Regge theory allows one to
write an unsubtracted dispersion relation in this case, leading to the
strictures\cite{DHG66}
\begin{eqnarray}
{\pi e^2\kappa^2\over 2M^2}&=& - \; \int_0^\infty {d\omega\over \omega}
[\sigma_-(\omega)-\sigma_+(\omega)] \nonumber\\
\gamma&=&{1\over 4\pi^2}\int_0^\infty{d\omega\over \omega^3}
[\sigma_-(\omega)-\sigma_+(\omega)]\label{eq:dhg}
\end{eqnarray}
where $\sigma_\pm(\omega)$ are the photo-absorption cross sections for
parallel and anti-parallel alignments of photon and target helicities.
The first of these relations---the
Drell-Hearn-Gerasimov (DHG) sum rule---has received a good deal of attention
recently.  On the experimental side, efforts are being mounted to
measure the spin-dependent structure function $f_2(\omega)$
directly, thereby confirming the prediction of the low energy
theorem.\cite{AS96}  However, this has not yet been achieved.  On the 
theoretical
end, there have been a number of attempts to evaluate the dispersive integral
of the DHG using what information currently exists for
the photo-absorption cross sections.  The existing data set is 
incomplete in that helicity-dependent cross sections have not yet been
measured.  Thus one uses existing multipole decompositions from
{\it unpolarized} experiments in order to perform the analysis. Such
decompositions are available, however, only in the single pion
production channel so that above the two pion threshold
model-dependent assumptions must be made.  With this caveat present
results are somewhat higher than predicted by the low energy theorem
\begin{eqnarray}
{\pi e^2\kappa_p^2\over 2M^2}&=&0.167{\rm GeV}^{-2}\nonumber\\
-\int_0^\infty \frac{d\omega}{\omega}[\sigma_-(\omega)-\sigma_+(\omega)]^{
\rm multipole}&=&\left\{
\begin{array}{ll}
0.208{\rm GeV}^{-2} & {\rm Karliner}\cite{K73}\\
0.210{\rm GeV}^{-2} & {\rm Workman/Arndt}\cite{WA92}\\
0.182{\rm GeV}^{-2} & {\rm Burkert/Li}\cite{BL93} 
\end{array}\right. \nonumber \\
& &
\end{eqnarray}
and there have even been arguments made that challenge the assumptions
under which the DHG form was derived.\cite{CLW94}  However, resolution of these
problems awaits reliable helicity-dependent cross section
measurements, which should be available in the near future.\footnote{A
second area of interest is in the generalization of the DHG form to
include the deep inelastic region and its connection with the 
integrated spin dependent structure function $g_1(k^2)$.  However, we
shall not discuss this issue herein.}

Using the same multipole analysis it is possible to evaluate the
dispersion integral involving the spin dependent polarizability,
yielding\cite{AMS94}
\begin{equation}
\gamma={1\over 4\pi^2}\int_0^\infty{d\omega\over \omega^3}
[\sigma_-(\omega)-\sigma_+(\omega)]^{\rm multipole}=\left\{
\begin{array}{ll}
-1.3\times 10^{-4}{\rm fm}^{-4}& p\\
-0.4\times 10^{-4}{\rm fm}^{-4}& n
\end{array}\right.\label{eq:MP}
\end{equation}
For a detailed discussion of the contributions of the various nucleon 
resonances
to the sum rules of Eq.(\ref{eq:dhg}) and the problems of the constituent
quark model in describing the DHG sum rule we refer the reader to the review
article by Drechsel \cite{D94}.

These results are in dramatic disagreement with the ${\cal O}(p^3)$ ChPT
predictions\cite{BKKM92}
\begin{equation}
\gamma_p=\gamma_n={e^2g_A^2\over 96\pi^3F_\pi^2m_\pi^2}=+ \; 4.4\times
10^{-4}{\rm fm}^{-4} \label{eq:gg3}
\end{equation}
Unlike the case of $\bar{\alpha}_E$ and $\bar{\beta}_M$, at present there
exists no corresponding ${\cal O}(p^4)$ ChPT calculation that reconciles this 
discrepancy.
Presently the best information about the sub-leading behaviour of $\gamma$
comes from a relativistic baryon ChPT 
calculation at the one-loop level \cite{BKKM92}, yielding
\begin{eqnarray}
\gamma_{p}^{\rm 1-loop} = + 2.2 \times 10^{-4} \; {\rm fm}^4 \;\;\;\; 
\gamma_{n}^{\rm 1-loop} = + 3.2 \times 10^{-4} \; {\rm fm}^4 \; . \label{eq:gr}
\end{eqnarray}
This calculation does not resolve the discrepancy with the multipole 
analysis results of Eq.(\ref{eq:MP}).
However, it is known from phenomenological considerations\footnote{This has
also been pointed out by the authors of \cite{BKKM92}.} that $\Delta(1232)$
makes a major contribution of opposite sign. In section 4.2 we present 
a systematic chiral calculation of $\gamma$ with nucleons, deltas and pions
as explicit degrees of freedom. 

Having given a brief summary of current research in this area we now
proceed to outline the formalism which allows inclusion of the
$\Delta$(1232) in a consistent chiral power counting framework.

\section{Small Scale Expansion in HBChPT}

The subject of SU(2) Heavy Baryon ChPT of nucleons and pions has been 
well developed in recent years \cite{BKKM92,BKM95}. In the 
conventional form
one organizes the calculation according to an ${\cal O}(p^n)$ powercounting,
where $p$ denotes either a soft momentum or the pion mass $m_\pi$. All 
nucleon resonances, strange particles, vector mesons, etc. are integrated
out, i.e. they only appear in higher order contact interactions. For near
threshold processes this program has proved highly successful--for an
outstanding review of the field we refer the reader to ref. \cite{BKM95}. 
However, phenomenologically it is known that in the nucleon sector the
first nucleon resonance--$\Delta$(1232)--lies very close to the nucleon 
and can exert its influence even to processes at very low energies. 
This special situation in the baryon sector has early on \cite{JM91} prompted 
suggestions to
keep the spin-3/2 baryon resonances as explicit degrees of freedom in the
chiral lagrangian in order to include all relevant physics and to improve
the convergence of the perturbation series.

If one retains $\Delta$(1232) as an explicit degree of freedom in SU(2) 
HBChPT, one is faced with an additional dimensionful 
parameter $\Delta = M_\Delta -M_N$, which corresponds to the mass splitting
between the nucleon and the delta resonance in the chiral limit. 
Phenomenologically $\Delta$ is
a small parameter of $\approx 300$ MeV which, unlike $m_\pi$ remains
finite in the chiral limit. As was shown in refs. \cite{JM91,HHK96,HHK96a} one
can nevertheless set up a consistent field-theoretic HBChPT formalism 
provided one organizes the calculation
according to an ${\cal O}(\epsilon^n)$ powercounting, where $\epsilon$ now
denotes a small scale of either a soft momentum, the pion mass or the 
mass splitting $\Delta$. Our formalism has been  constructed in 
such a way, that an ${\cal O}(\epsilon^n)$ result {\em automatically} contains
any ${\cal O}(p^n)$ result in \cite{BKM95} plus additional 
terms involving the $\Delta$(1232) resonance which might be present to
the order we are calculating.

In our calculation of Compton scattering below, which is done 
to ${\cal O}(\epsilon^3)$ in the small scale expansion, we
shall need only the lowest order vertices, the propagator involving the
$\Delta(1232)$, as well as possible counterterm contributions to $NN\gamma$,
$NN\gamma\gamma$, and $N\Delta\gamma$.  Details of how this can be achieved 
in a general $1/M$ expansion are given in ref. \cite{HHK96,HHK96a}.  Here 
we list only the minimal results necessary for the present calculation.

The systematic 1/M-expansion of the coupled $N\Delta$-system starts with 
the most general chiral invariant lagrangian 
involving relativistic spin 1/2 ($\psi_N$) and spin 3/2 ($\psi_\mu$) fields
\footnote{In order to take into account the isospin 3/2 property of 
$\Delta$(1232) we
supply the Rarita-Schwinger spinor with an additional isospin index $i$, 
subject to the subsidiary condition $\tau^i \; \psi_{\mu}^i (x) = 0$.}
\begin{equation}
{\cal L}={\cal L}_N + {\cal L}_\Delta + \left( {\cal L}_{\Delta N} + h.c.
\right) \label{eq:4L}
\end{equation}
with 
\begin{eqnarray}
{\cal L}_{\Delta} & = & \bar{\psi}^{\mu}_i \; \Lambda_{\mu\nu}^{ij} \; 
\psi^{\nu}_j \nonumber \\
\Lambda_{\mu\nu}^{ij} & = & - \mbox{\Large [} ( i \not\!\!{D}^{ij} - M_{\Delta}
                            \; \delta^{ij} ) g_{\mu \nu} - \frac{1}{4}
                            \gamma_{\mu} \gamma^{\lambda} ( i \not\!\!{D}^{ij}
                            - M_{\Delta} \; \delta^{ij}) \gamma_{\lambda}
                            \gamma_{\nu} \nonumber \\
                      &   & \; \; \; + \frac{g_{1}}{2} g_{\mu \nu}
                            \not\!{u}^{ij}
                            \gamma_{5} + \frac{g_{2}}{2} ( \gamma_{\mu}
                            u_{\nu}^{ij} + u_{\mu}^{ij} \gamma_{\nu} )
                            \gamma_{5} + \frac{g_{3}}{2} \gamma_{\mu}
                            \not\!{u}^{ij} \gamma_{5} \gamma_{\nu}
                            \mbox{\Large ]} \label{eq:Lambda} + ...
\nonumber \\
{\cal L}_N &=& \bar\psi_N \left(i\not\!\!{D}-M_N+{g_A\over 2}\not\!{u} \gamma_5
\right) \psi_N + ... \nonumber \\
{\cal L}_{\Delta N} &=& g_{\pi N \Delta} \; \bar{\psi}^{\mu}_{i} \left(
g_{\mu\nu}+z \gamma_\mu \gamma_\nu \right) w^{\nu}_{i} \psi_N + ...
\end{eqnarray}
We have only displayed the leading order terms, the dots denote terms with 
more derivatives or insertions of the 
light quark mass matrix. It is understood that they are included up to the order
one is working in the $\epsilon$-expansion.
Following the conventions of SU(2) HBChPT in the spin 1/2 sector
\cite{GSS88,BKM95}, we have defined the following structures
\begin{eqnarray}
D_{\mu}^{ij} \; \psi^{\nu}_j & = & \left( \partial_{\mu} \; \delta^{ij} +
                                   \Gamma_{\mu}^{ij} \right) \psi^{\nu}_j
                                   \nonumber \\
\Gamma_{\mu}^{ij} & = & \Gamma_\mu \; \delta^{ij} - \frac{i}{2} \;
                        \epsilon^{ijk} \; Tr [ \tau^k \; \Gamma_\mu ]
                        \nonumber \\
\Gamma_{\mu}       & = & \frac{1}{2} \left[ u^{\dag} , \partial_{\mu} u \right]
                         - \frac{i}{2} u^{\dag} ( {\bf v}_{\mu} +
                         {\bf a}_{\mu} ) u - \frac{i}{2} u ( {\bf v}_{\mu} -
                         {\bf a}_{\mu} ) u^{\dag}
                         \nonumber \\
u_\mu^{ij}         &=&  u_\mu\delta^{ij}-i\epsilon^{ijk}w_\mu^k \nonumber\\
w_\mu^i            &=& \frac{1}{2} Tr
                         \left[ \tau^i u_\mu \right] \nonumber \\
u_{\mu}            & = & i u^{\dagger} \nabla_{\mu} U u^{\dagger}\nonumber\\
\nabla_{\mu} U     & = & \partial_{\mu} U - i ( {\bf v}_{\mu} + {\bf a}_{\mu}
                         ) U + i U ( {\bf v}_{\mu} - {\bf a}_{\mu} )
                         \nonumber \\
U                  & = & u^{2} = exp \left( \frac{i}{F_{\pi}} {\bf \vec{\tau}
                         \cdot \vec{\pi} } \right) \; ,
\label{eq:definitions}
\end{eqnarray}
where ${\bf v}_\mu$, ${\bf a}_\mu$ denote external vector, axial-vector 
fields.

The ``light'' degrees of freedom of the spin 3/2 field, which are retained 
in the effective low-energy theory, are identified as
\begin{equation}
T_{\mu}^i (x) \equiv  P_{v}^{+} \; P^{3/2}_{(33)\mu\nu} \; \psi^{\nu}_i (x)
                    \; \mbox{exp}(i M v \cdot x) \label{eq:T}
\end{equation}
where we have introduced a spin 3/2 projection operator for fields with 
{\it fixed velocity} $v_\mu$
\begin{equation}
P^{3/2}_{(33)\mu \nu}  =  g_{\mu \nu} - \frac{1}{3} \gamma_{\mu} \gamma_{
                               \nu} - \frac{1}{3} \left( \not\!{v} \gamma_{\mu}
                               v_{\nu} + v_{\mu} \gamma_{\nu}\not\!{v} 
\right) . \label{eq:proj}
\end{equation}
The remaining components,
\begin{equation}
G_{\mu}^i (x)  =  \left( g_{\mu\nu}-P_{v}^{+} \; P^{3/2}_{(33)\mu\nu}\right)
                   \psi^{\nu}_i (x) \; \mbox{exp}(i M v \cdot x) ,
\label{eq:G}
\end{equation}
are shown to be ``heavy'' \cite{HHK96,HHK96a} and are integrated out.

Rewriting the lagrangians of Eq.(\ref{eq:4L}) in terms of the spin 3/2
heavy baryon components $T_{\mu}$ and $G_{\mu}$, and the corresponding
``light'' and ``heavy'' spin 1/2 components $N$, $h$, defined as
\begin{eqnarray}
N(x)&=& P_v^+ \; \psi_N \; {\rm exp} (i M v\cdot x) \nonumber \\
h(x)&=& P_v^- \; \psi_N \;  {\rm exp} (i M v\cdot x),
\label{heavyN}
\end{eqnarray}
the general lagrangians take the form
\begin{eqnarray}
L_{N} &=& \bar{N} {\cal A}_{N} N + \left( \bar{h} {\cal B}_{N} N + h.c. \right)
          - \bar{h} {\cal C}_{N} h \nonumber \\
L_{\Delta N} &=& \bar{T} {\cal A}_{\Delta N} N + \bar{G} {\cal B}_{\Delta N} N +
                 \bar{h} {\cal D}_{N \Delta} T + \bar h {\cal C}_{N \Delta} G +
                  h.c. \nonumber\\
L_{\Delta}&=&\bar{T} {\cal A}_{\Delta} T + \left( \bar{G} {\cal B}_{\Delta} T +
             h.c. \right) - \bar{G} {\cal C}_{\Delta} G.
\label{Lgeneral}
\end{eqnarray}
Matrices ${\cal A}$, ${\cal B}$, ${\cal C}$ admit a low energy scale 
expansion--${\cal A}$, ${\cal B}$ start at order $\epsilon^1$, whereas 
${\cal C}$ has 
a leading term of order 1. This allows to perform a systematic 
1/M-expansion, following an approach developed by Mannel et al. in the field
of heavy quark physics 
\cite{MRR92} , which was later applied to spin 1/2 HBChPT by Bernard et al.
\cite{BKKM92}. The result of this procedure is the effective action for 
the coupled $N\Delta$-system \cite{HHK96,HHK96a}
\begin{equation}
S_{\rm eff}= \int d^4x \left\{ \bar T \tilde {\cal A}_{\Delta} T
+\bar N \tilde {\cal A}_{N} N
+\left[ \bar T \tilde {\cal A}_{\Delta N} N + h.c.\right] \right\}
\label{Seff}
\end{equation}
with
\begin{eqnarray}
\tilde {\cal A}_\Delta &=& {\cal A}_\Delta
+ \gamma_0 \tilde {\cal D}_{N \Delta}^\dagger \gamma_0 \tilde {\cal C}_N^{-1}
\tilde {\cal D}_{N \Delta}
+ \gamma_0 {\cal B}_\Delta^\dagger \gamma_0 {\cal C}_\Delta^{-1} {\cal B}_\Delta
\nonumber \\
\tilde {\cal A}_N &=& {\cal A}_N
+ \gamma_0 \tilde {\cal B}_{N}^\dagger \gamma_0 \tilde {\cal C}_N^{-1}
\tilde {\cal B}_{N}
+ \gamma_0 {\cal B}_{\Delta N}^\dagger \gamma_0 {\cal C}_\Delta^{-1}
{\cal B}_{\Delta N}
\nonumber \\
\tilde {\cal A}_{\Delta N} &=& {\cal A}_{\Delta N}
+ \gamma_0 \tilde {\cal D}_{N \Delta}^\dagger \gamma_0 \tilde {\cal C}_N^{-1}
\tilde {\cal B}_{N}
+ \gamma_0 {\cal B}_\Delta^\dagger \gamma_0 {\cal C}_\Delta^{-1}
{\cal B}_{\Delta N}
\label{eq:Atilde}
\end{eqnarray}
and
\begin{eqnarray}
\tilde {\cal C}_N&=& {\cal C}_N-{\cal C}_{N \Delta}
{\cal C}_{\Delta}^{-1} \gamma_0 {\cal C}_{N \Delta}^\dagger \gamma_0
\nonumber \\
\tilde {\cal B}_N&=& {\cal B}_N+{\cal C}_{N \Delta }
{\cal C}_{\Delta}^{-1} {\cal B}_{\Delta N} \nonumber \\
\tilde {\cal D}_{N\Delta}&=& {\cal D}_{N \Delta}
+ {\cal C}_{N \Delta} {\cal C}_{\Delta}^{-1} {\cal B}_{\Delta} .
\label{BCtilde}
\end{eqnarray}

The vertices relevant for our calculation can be read off directly from
Eq.(\ref{Seff}). Further analysis below will be divided into two classes of 
contributions, one-loop graphs and Born graphs. As one can see from 
Eq.(\ref{Seff}), any calculation at ${\cal O}(\epsilon^3)$ which has a nucleon
in the initial (and final) state, {\em completely}
contains a ${\cal O}(p^3)$ calculation in the formalism of \cite{BKKM92}. 
Furthermore, it shows unambiguously, which additional diagrams involving 
$\Delta$(1232) must be appended.

\subsection{Loop Graphs}

To order $\epsilon^3$, only one-loop graphs with vertices of order $\epsilon$
must be considered. Thus,
the pieces in Eq.(\ref{Seff}) we need are ${\cal A}_N^{(1)}$, 
${\cal A}_\Delta^{(1)}$ and ${\cal A}_{\Delta N}^{(1)}$. More explicitly, 
these are found as
\begin{eqnarray}
{\cal A}_N^{(1)} &=& iv\cdot D +g_A \; S\cdot u \nonumber \\
{\cal A}_{\Delta N}^{(1)}  &=& g_{\pi N\Delta} w_{\mu}^i \nonumber\\
{\cal A}_{\Delta}^{(1)} &=& - \left[ i \; v \cdot D^{ij} - \Delta \;
\delta^{ij} + g_{1} \; S \cdot u^{ij} \right] \; g_{\mu \nu}
\label{eq:llead}
\end{eqnarray}
where $S_\mu$ denotes the Pauli-Lubanski spin vector \cite{BKM95}. From 
matrices
${\cal A}_N^{(1)}$ and  ${\cal A}_\Delta^{(1)}$ 
we determine the SU(2) HBChPT propagators in momentum space with soft 
momentum\footnote{Working to ${\cal O}(\epsilon^3)$ and in the absence of 
any ${\cal O}(\epsilon)$ Born diagrams due to our choice of gauge, we always
can identify the large mass $M$ with the physical mass of the nucleon $M_N$.
Mass renormalization only becomes important in an ${\cal O}(\epsilon^4)$
HBChPT calculation of Compton scattering.}
$r_\mu = p_\mu - M v_\mu$:
\begin{eqnarray}
S^{1/2} ( v \cdot r ) &=& \frac{i}{v \cdot r + i\eta} \nonumber \\
S_{\mu\nu}^{3/2} ( v \cdot r ) &=& \frac{- \; i\; P^{3/2}_{\mu\nu}}{v \cdot r
                                   - \Delta + i\eta} \; \xi^{ij}_{I=3/2} \; ,
\label{eq:propagator}
\end{eqnarray}
with $P^{3/2}_{\mu\nu}$ denoting the spin 3/2 HBChPT projector in 
d-dimensions \cite{HHK96a} 
\begin{equation}
P^{3/2}_{\mu\nu} = g_{\mu\nu} - v_{\mu\nu} + \; \frac{4}{d-1} \; S_\mu S_\nu
\end{equation}
and 
\begin{equation}
\xi^{ij}_{I=3/2} = \delta^{ij} - \frac{1}{3} \; \tau^i \tau^j
\end{equation}
being the corresponding isospin 3/2 projector. From 
Eq.(\ref{eq:propagator}) one can
see that the delta propagator counts as a quantity of order $\epsilon^{-1}$ 
in our expansion scheme.

At present, we have no systematic determination of the coupling constant 
$g_{\pi N\Delta}$ within the small scale expansion. For the time being 
we rely on the phenomenological analysis of Davidson, Mukhopadhyay and Wittman
(DMW) \cite{DMW91}, yielding
\begin{equation}
g_{\pi N\Delta}^{HHK} = \frac{F_\pi}{m_\pi} \; g_{\pi N\Delta}^{DMW}
                        \approx 1.5 \; \pm 0.2 \; . \label{eq:gpnd}
\end{equation}
Finally,
choosing the velocity vector $v_\mu = (1,0,0,0)$ and working in the Coulomb
gauge $v \cdot \epsilon =v \cdot \epsilon^\prime = 0$, 
we conclude that we have to calculate 18 loop diagrams,
displayed in Fig. 3 and Fig. 4. Details of the calculation are given
in Appendices B and C, the results will be discussed in section 4.
\newpage

\subsection{Born graphs}

\begin{table}
\begin{center}
\begin{tabular}{l|l}
vertex & lagrangian \\ \hline
O($\epsilon$) $\gamma NN$ & ${\cal A}_{N}^{(1)} \rightarrow 0$ in Coulomb 
gauge \\
O($\epsilon$) $\gamma N\Delta$ & $--$ \\
O($\epsilon^2$) $\gamma NN$ & ${\cal A}_N^{(2)}$ and 
$\gamma_0 {\cal B}_N^{(1) \; \dagger} \gamma_0 
{\cal C}_N^{(0) \; -1} {\cal B}_N^{(1)}$ \\
O($\epsilon^2$) $\gamma\Delta N$ & ${\cal A}_{\Delta N}^{(2)}$ \\ 
O($\epsilon^2$) $\gamma\gamma NN$ & $\gamma_0 {\cal B}_N^{(1) \; \dagger} 
\gamma_0 {\cal C}_N^{(0) \; -1} {\cal B}_N^{(1)}$ \\
O($\epsilon^3$) $\gamma\gamma NN$ & $\gamma_0 {\cal B}_N^{(2) \; \dagger} 
\gamma_0 {\cal C}_N^{(0) \; -1} {\cal B}_N^{(1)}$ + h.c. 
\end{tabular} \newline
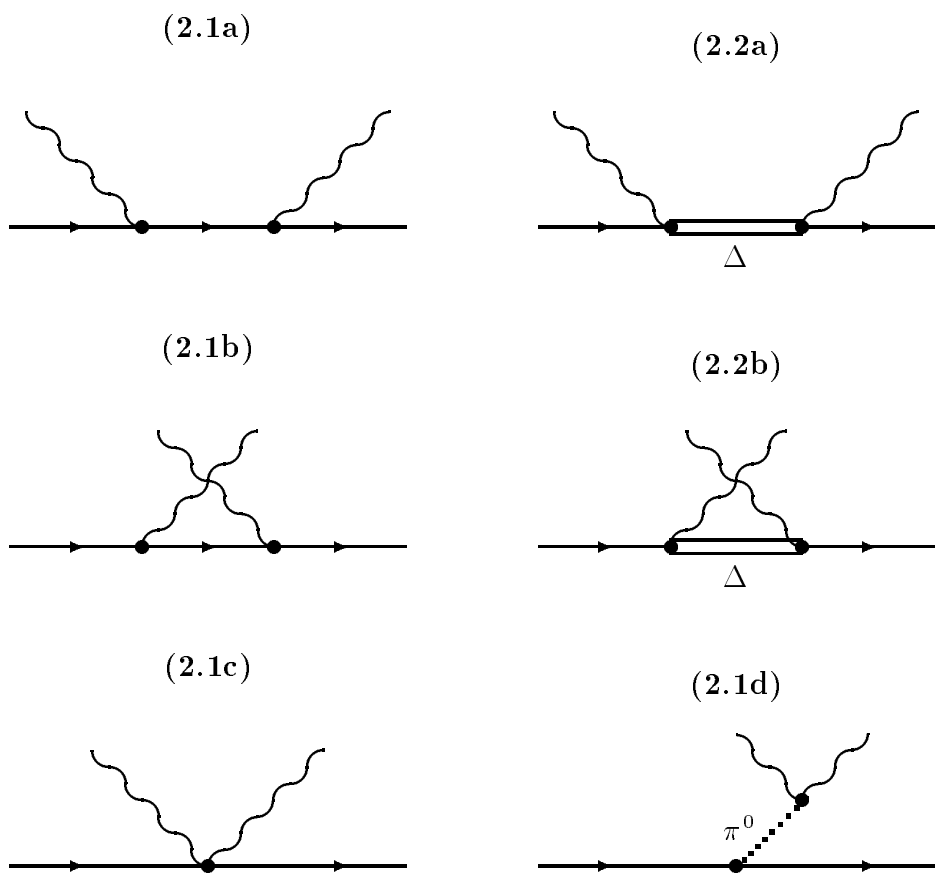
\caption{Vertices for Born graphs in Forward Compton Scattering}
\end{center}
\end{table}

The Born graphs contributing at ${\cal O}(\epsilon^3)$ Compton scattering 
are shown in Figs. 1 and 2,
with the pertinent vertices given in Table 1. The structures not involving 
the delta resonance  can be taken from 
ref. \cite{Ecker96}\footnote{In general this is not the case. As will be shown
in \cite{HHK96a} the ``heavy components'' of the relativistic spin
3/2 field modify the counterterms of the $NN$ lagrangian starting at 
${\cal O}(\epsilon^2)$. In our specific case the $NN$ vertices are unchanged.
}. We note that our
use of the Coulomb gauge dramatically reduces the number of diagrams.
Also, the possible diagram of Fig. 2.1d with an 
anomalous $\pi^0 \rightarrow \gamma\gamma$ vertex does not contribute in 
forward direction. 
Due to the fact that the photo-excitations of $\Delta$(1232) begin with 
the M1 transition, there exists no $\gamma N\Delta$ vertex at 
${\cal O}(\epsilon)$. Consequently, there is no 1/M-corrected interaction
of this type at 
${\cal O}(\epsilon^2)$ in Table 1. However, the relativistic 
counterterm lagrangian \cite{HHK96,DMW91}
\begin{equation}
{\cal L}_{c.t.}^{N\Delta}=\frac{i b_1}{2M_N} \bar\psi^{\mu}_i 
\left( g_{\mu\nu} + y \gamma_\mu \gamma_\nu \right) \gamma_\rho \gamma_5
\; \frac{1}{2} \; Tr \left[ f_{+}^{\rho\nu} \; \tau^i \right] \psi_N, 
\label{eq:b1rel}
\end{equation}
provides the M1 $\gamma N\Delta$ transition strength and leads
to an ${\cal O}(\epsilon^2)$ structure:
\begin{equation}
{\cal A}_{\Delta N}^{(2)}={i b_1 \over M_N} S_\nu \; \frac{1}{2} \; Tr \left[ 
f_+^{\nu\lambda} \; \tau^i \right],
\label{gammavertex}
\end{equation}
with the chiral field tensor being defined as \cite{BKM95}
\begin{eqnarray}
f_{\mu\nu}^\pm = u^\dagger \; F_{\mu\nu}^R \; u \; \pm \; u \; F_{\mu\nu}^L
                 \; u^\dagger \nonumber \\
F_{\mu\nu}^{L,R} = \partial_\mu F_{\nu}^{L,R} - \partial_\nu F_{\mu}^{L,R}
                   -i \; \left[ \; F_{\mu}^{L,R} \; , \; F_{\nu}^{L,R} \;
                   \right] \nonumber \\
F_{\mu}^R = {\bf v_\mu} + {\bf a_\mu} , \; \; F_{\mu}^L = {\bf v_\mu} - {
            \bf a_\mu} \; .
\end{eqnarray}
We note that the off-shell parameter $y$ in Eq.(\ref{eq:b1rel}) does not 
contribute at order $\epsilon^2$ when going to the effective heavy baryon 
lagrangian--it will only enter at ${\cal O}(\epsilon^3)$. However, since the
Delta propagator Eq.(\ref{eq:propagator}) counts as order $\epsilon^{-1}$, 
and we effectively have no ${\cal O}(\epsilon)$ vertices around, we do not
have to consider $\epsilon^3$ $\gamma NN$ or $\gamma N\Delta$ vertices to 
the order we are working. Furthermore, the absence of an ${\cal O}(\epsilon)$ 
$\gamma N\Delta$ vertex is also 
responsible for the fact that the ${\cal O}(\epsilon^3)$ two-photon seagull
term does not get renormalized by Delta interactions. The only Born diagrams
involving $\Delta$(1232) are therefore s- and u-channel resonance exchange
with vertices from ${\cal A}_{\Delta N}^{(2)}$ (diagrams 2.2a,b in Figure 2).
At present, the magnitude of the finite ${\cal O}(\epsilon^2)$ counterterm 
$b_1$ is not known very accurately. Until one has identified a suitable 
$b_1$ dependent observable which has both been calculated in the small scale 
expansion and been reasonably well measured\footnote{Work along these lines
is under way; V. Bernard, T.R. Hemmert, J. Kambor and U.-G. Mei\ss ner, in
preparation.}, we will employ a phenomenological 
relation found in ref. \cite{DMW91}, which in our convention reads  
\begin{equation}
b_1 \approx - \; 2.3 \; \frac{m_\pi}{2 F_\pi} \; g_{\pi N\Delta}^{HHK} \approx
- \; ( \; 2.5 \; \pm \; 0.35 \; ).
\end{equation} 
For the numerical estimate we have used $g_{\pi N\Delta}^{HHK}$ of 
Eq.(\ref{eq:gpnd}). This completes the background necessary for the 
present application.

\section{Forward Compton Scattering and Nucleon Polarizabilities}

Having outlined our formalism, in this section we 
investigate the influence of the $\Delta (1232)$
resonance on the nucleon polarizabilities. We restrict ourselves
to the case of forward scattering, which provides information on the electric
polarizability $\bar{\alpha}_E$, the magnetic polarizability
$\bar{\beta}_M$ and the spin 
polarizability $\gamma$. As mentioned above, 
for the spin-averaged quantities $\bar{\alpha}_E$ and $\bar{\beta}_M$
the results involving only nucleon and pion degrees of freedom 
are already known to ${\cal O}(p^{4})$ 
\cite{BKSM93} in the chiral expansion.  Here we present
a systematic analysis of the $\Delta(1232)$ contributions to $\bar{\alpha}_E$, 
$\bar{\beta}_M$ and
$\gamma$ at ${\cal O}(\epsilon^{3})$. A complete calculation of all polarizabilities at 
${\cal O}(\epsilon^{4})$ including pion, nucleon and Delta
degrees of freedom will be the subject of future work.

\subsection{Spin-averaged Forward Compton Scattering}

Working in the gauge---$\epsilon\cdot v=0$---the spin-averaged Compton 
tensor in forward direction $\Theta_{\mu\nu}$ can be written as
\begin{eqnarray}
\epsilon^{\prime \;\mu} \; \Theta_{\mu\nu} \; \epsilon^{\nu} & = & 
                  e^{2} \; \epsilon^{\prime \; \mu} \epsilon^{\nu} \frac{1}{2}
                  Tr \left[ \; P_{v}^{+} \; T_{\mu\nu}(v,k) \; \right] 
                  \nonumber \\
            & = & e^{2} \left[ \; \epsilon^\prime \cdot \epsilon \; 
                  U(\omega) \; + \; \epsilon^\prime \cdot k \; \epsilon \cdot 
                  k \; V(\omega) \right] \;, \label{eq:Amp}
\end{eqnarray}
where $k$ is the four-momentum of a photon with energy $\omega = k \cdot v$.
$\epsilon \; (\epsilon^\prime)$ refer to the polarization vector of the 
incoming (outgoing) forward scattering photon and $T_{\mu\nu}(v,k)$ is the 
Fourier-transformed matrix element of two time-ordered electromagnetic currents
\begin{equation}
T_{\mu\nu}(v,k) = \int d^{4}x \; e^{i k \cdot x} \langle N(v) | \mbox{\em T}
\left[ \; J_{\mu}^{em}(x) J_{\nu}^{em}(x) \;\right] | N(v) \rangle
\end{equation}
In the spin-averaged case all the information about the low-energy structure
of the photon is contained in just two functions\footnote{The auxiliary 
function 
$V(\omega)$ can be eliminated for real photons. Nevertheless one can obtain 
information about the magnetic polarizability $\beta$ from it.
}-- $U(\omega),V(\omega)$. However, there exists a structure-independent 
constraint \cite{GMGL}
with respect to $U(\omega)$, stating that in the limit of zero photon energy
one has to obtain the Thomson result of Eq.(\ref{eq:1})
\begin{equation}
U(0) = Z^2 / M \; , \label{eq:TL}
\end{equation}
where $Z$ refers to the charge number of the Compton target and $M$ is its
mass. Furthermore, $U(\omega)$ has to be even under crossing 
symmetry\footnote{$U(\omega)$ is related to the function $f_1(\omega)$ of 
Eq.(\ref{eq:f1f2}) via $f_1(\omega)= - \frac{e^2}{4\pi} \; U(\omega)$.},
{\em i.e.} $U(\omega)=U(- \omega)$. 

Keeping these two non-trivial constraints in
mind we split up the calculation of $U(\omega),\;V(\omega)$ into four separate
components:
\begin{eqnarray}
U(\omega) & = & U_{N}(\omega)^{Born} + U_{N}(\omega)^{loops} + 
                U_{\Delta}(\omega)^{Born} + U_{\Delta}(\omega)^{loops} 
                \nonumber \\
V(\omega) & = & V_{N}(\omega)^{Born} + V_{N}(\omega)^{loops} + 
                V_{\Delta}(\omega)^{Born} + V_{\Delta}(\omega)^{loops} 
                \label{eq:U4}
\end{eqnarray}
We start with the calculation of the nucleon Born contributions to $U(\omega),
\; V(\omega)$. In the Coulomb gauge $\epsilon \cdot v = 0$ there exist non-zero
contributions at ${\cal O}(\epsilon^2)$ and ${\cal O}(\epsilon^3)$, as shown
by the diagrams in Figures 1 and 2. To ${\cal O}(\epsilon^3)$ we find
\begin{eqnarray}
U_{N}(\omega)^{Born} & = & \frac{1}{M_{N}} \; \frac{1}{2} (1 + \tau_{3}) 
                           \nonumber \\
V_{N}(\omega)^{Born} & = & \frac{1}{M_{N}^{2} \omega} \; \frac{1}{2} (1 + 
                           \tau_{3}),
\end{eqnarray}
with $U_N(\omega)$ solely stemming from the ${\cal O}(\epsilon^2)$ seagull
diagram of Figure 1 and $V_N(\omega)$ arising from diagrams 2.1a and 2.1b
of Figure 2. We also
note that $U_N (0)^{Born}$ satisfies the Thomson limit Eq.(\ref{eq:TL})
for proton ($Z=1$) and neutron ($Z=0$) targets, as expected. We therefore
conclude that
\begin{eqnarray}
U_{N}(0)^{loops}  = U_{\Delta}(0)^{Born} = U_{\Delta}(0)^{loops}  =  0 
\label{eq:Tnpi} \; ,
\end{eqnarray}
and this will serve as a powerful constraint and check on our calculation.

The ${\cal O}(\epsilon^{3})$ nucleon loop contributions to the spin-averaged 
functions can be 
obtained from the nine diagrams shown in Fig. 3 yielding\footnote{ 
Eq.(\ref{eq:UN}) agrees with the ${\cal O}(p^3)$ result of 
\cite{BKKM92}.}, as detailed in Appendix B
\begin{eqnarray}
U_{N}(\omega)^{loops} & = & - \; {11g_A^2\omega^2\over 192\pi F_\pi^2m_\pi}
                            + {\cal O}(\omega^4) \nonumber\\
V_{N}(\omega)^{loops} & = & - \; {g_A^2\over 192\pi F_\pi^2m_\pi} + 
                            {\cal O}(\omega^2) \label{eq:UN}
\end{eqnarray} 
Note that the loop effects are isospin-independent ({\it i.e.}
identical for neutron and proton)
to this order. We have also checked that 
$U_{N}(\omega=0)^{loops}=0$ (Eq.(\ref{eq:Tnpi})), by carefully analyzing the
dimensionality dependence of the nine loop amplitudes as given in Appendix
B.

Next, we evaluate $\Delta$(1232) Born contributions to $U(\omega), \; 
V(\omega)$. At
${\cal O}(\epsilon^3)$ we find two contributing diagrams, as shown in Fig. 2.
They yield
\begin{eqnarray}
U_\Delta(\omega)^{Born} &=& - \; \frac{8 b_1^2 \omega^2}{9 M_{N}^2}
                            {\Delta\over \Delta^2-\omega^2} \nonumber \\
V_\Delta(\omega)^{Born} &=& - \; \frac{8 b_1^2}{9 M_{N}^2}
                            {\Delta\over \Delta^2-\omega^2} 
\end{eqnarray}
where $b_1$ is the M1 $\Delta N\gamma$ coupling of Eq.(\ref{eq:b1rel}). Again,
 the Thomson constraint
$U_\Delta (0)^{Born}=0$ holds, as required by Eq.(\ref{eq:Tnpi}).

Finally, the corresponding delta loop contributions can be found from the nine 
diagrams shown in Figure 4. Analyzing the invariant amplitudes given in
Appendix C, we find
\begin{eqnarray}
U_\Delta (\omega)^{loop}&=& - \; \frac{g_{\pi N\Delta}^2 \omega^2}{54\pi^2 
                            F_{\pi}^2} \left[ \frac{9 \Delta}{\Delta^2 - m_{
                            \pi}^2} - \frac{9 m_{\pi}^2}{(\Delta^2 - m_{
                            \pi}^2)^{3/2}} \log R + \frac{2}{\sqrt{\Delta^2 -
                            m_{\pi}^2}} \log R \right] \nonumber \\
                        & & + \; {\cal O}(\omega^4) \nonumber\\
                        & & \nonumber \\
V_\Delta(\omega)^{loop} &=& - \; \frac{g_{\pi N\Delta}^2}{54\pi^2 F_{\pi}^2}
                            \frac{1}{\sqrt{\Delta^2 - m_{\pi}^2}} \log R
                            + \; {\cal O}(\omega^2) \; ,
\end{eqnarray}
with $R$ defined as
\begin{equation}
R=\frac{\Delta}{m_\pi} + \sqrt{ \frac{\Delta^2}{m_{\pi}^2} -1} .\label{eq:R}
\end{equation}
Again we note the validity of the Thomson stricture
$U_{\Delta}(\omega=0)^{loops}=0$. 

In order to extract the nucleon electric and magnetic 
polarizabilities, we now define the nucleon Born
term subtracted quantities $\hat{U}(\omega), \hat{V}(\omega)$
\begin{eqnarray}
\hat{U}(\omega) & = & U(\omega) - U_{N}(\omega)^{Born} \\
\hat{V}(\omega) & = & V(\omega) - V_{N}(\omega)^{Born} 
\end{eqnarray}
and make the connections
\begin{eqnarray}
\bar{\alpha}_E + \bar{\beta}_M & = & - \; \frac{e^{2}}{8\pi} \; \frac{
                                     \partial^{2}}{\partial \omega \; ^{2}} \;
                                     \hat{U}(\omega) |_{\omega=0} \\
                 \bar{\beta}_M & = & - \; \frac{e^{2}}{4\pi} \; \hat{V}(
                                     \omega=0)
\end{eqnarray}
\newpage

Adding up all three contributions\footnote{The $N\pi$-loop parts of 
Eqs.(\ref{eq:alpha},\ref{eq:beta}) agree with the ${\cal O}(p^3)$ calculation
of \cite{BKKM92}. In SU(3) an estimate of spin 3/2 contributions to 
$\bar{\alpha}_E$ and $\bar{\beta}_M$ was given in \cite{BS92}. 
$\bar{\alpha}_E$ and $\bar{\beta}_M$ have also been calculated in
SU(2) relativistic baryon ChPT \cite{BKM91}.}  
one finds $\bar{\alpha}_E$ and 
$\bar{\beta}_M$ to ${\cal O}(\epsilon^3)$:
\begin{eqnarray}
\bar{\alpha}_E 
       & = & + \; \frac{e^2}{4 \pi} \;\frac{5 g_{A}^{2}}{96 \pi F_{\pi}^{2}}
             \; \frac{1}{m_{\pi}} \nonumber \\
       &   & \nonumber \\
       &   & + \; 0 \nonumber \\
       &   & \nonumber \\
       &   & + \; \frac{e^2}{4 \pi} \;\frac{g_{\pi N\Delta}^2}{54 \pi F_{
             \pi}^2} \; \frac{1}{\pi} \left[ \; \frac{9 \Delta}{\Delta^2 -
             m_{\pi}^2} + \ \frac{\Delta^2 - 10 m_{\pi}^2}{(\Delta^2 -
             m_{\pi}^2 )^{3/2}} \; \log R \; \right] \label{eq:alpha} \\
       &   & \nonumber \\
       & = & \left[12.2 \; \mbox{(N-loop)} + \; 0 \; \mbox{(delta-pole)} + 
             \; 8.6 \; \mbox{(delta-loop)} \right] \; \times 10^{-4} \; 
             {\rm fm}^3 
             \nonumber  \\
\bar{\beta}_M  
       & = & + \; \frac{e^2}{4\pi} \; \frac{g_{A}^2}{192 \pi F_{\pi}^2} 
             \; \frac{1}{m_\pi} \nonumber \\
       &   & + \; \frac{e^2}{4\pi} \; \frac{8 \; b_{1}^2}{9 \; M_{N}^2} \;
             \frac{1}{\Delta} \nonumber \\
       &   & + \; \frac{e^2}{4\pi} \;\frac{g_{\pi N\Delta}^2}{54 \pi F_{\pi}^2}
             \; \frac{1}{\sqrt{\Delta^2 - m_{\pi}^2}} \; \frac{1}{\pi}
             \log R \label{eq:beta} \\
       & = & \left[1.2 \; \mbox{(N-loop)} + \; 12 \; \mbox{(delta-pole)} + 
             \; 1.5 \; \mbox{(delta-loop)} \right] \; \times 10^{-4} \; 
             {\rm fm}^3 \nonumber 
\end{eqnarray}
In assessing these results, we observe that from the delta pole terms there 
exists a significant
contribution to the magnetic polarizability but none to the
corresponding electric polarizability at this order ! The strong effect
on $\bar{\beta}_M$ is not
surprising, as the large M1 nucleon-delta coupling is known to contribute
substantially to the magnetic polarizability \cite{PS95,M9}.
Also, we note that at ${\cal O}(\epsilon^3)$ the $\Delta\pi$-loop 
contributions to $\bar{\beta}_M$ are of the same size as the $N\pi$-loop
effects, which is unexpected. Furthermore, and perhaps most surprisingly, 
we find a large $\Delta\pi$-loop
component in $\bar{\alpha}_E$ ! Even though the
numerical values of the delta contributions in 
Eqs.(\ref{eq:alpha},\ref{eq:beta}) are understood to have sizable error bars
due to quadratic dependence on the presently poorly known couplings 
$g_{\pi N\Delta}$ and $b_1$, it is clear that one cannot expect to achieve
agreement with experiment at this order of the calculation--the effects
of $\Delta$(1232) are sizable and strongly renormalize the $N\pi$-loop results.
In the ${\cal O}(p^4)$ calculation of \cite{BKSM93} it has been 
shown that the
large delta pole-contribution in $\bar{\beta}_M$ cancels to a large extent 
against $N\pi$-loop
effects at that order. In addition to the simple delta poles the counterterms
of BKSM have been determined via 
``resonance saturation''. The large $\Delta\pi$-loop effects that we find
at ${\cal O}(\epsilon^3)$ are therefore not accounted for\footnote{In 
particular, the 
counterterm contribution to $\bar{\alpha}_E$ has been estimated to be
$\delta \alpha \approx 2.0 \times 10^{-4} \; {\rm fm}^3$ \cite{BKSM93} !} in 
the counterterms
of \cite{BKSM93} ! Unless one finds a cancelation through $\Delta\pi$-loops
in a future ${\cal O}(\epsilon^4)$ calculation, one would have to conclude that
``resonance saturation'' via simple pole-graphs is highly suspect  for 
counterterms in the baryon sector\footnote{Of course these counterterms can
still be determined from other experiments. A breakdown of the 
``resonance saturation'' hypothesis only means that the numerical value of 
the counterterms cannot be understood within a simple resonance {\em model}.}.
Future work will address this important question. Finally, we note that the
results for $\bar{\alpha}_E$ and $\bar{\beta}_M$ are isospin independent
at ${\cal O}(\epsilon^3)$. The surprising experimental results 
$\bar{\alpha}_{E}^n > \bar{\alpha}_{E}^p$, $\bar{\beta}_{E}^n > \bar{
\beta}_{E}^p$ cannot be addressed at this order and also warrant an
investigation at sub-leading order.

In order to obtain an estimate of the convergence of the perturbation
series and to check the field-theoretic consistency of our calculation, it is
useful to perform
a chiral expansion of our results for the polarizabilities, yielding
\begin{eqnarray}
\bar{\alpha}^{\; \chi}_E & = & + \; \frac{e^2}{4\pi} \; \frac{1}{6\pi 
                               F_{\pi}^2} \; \frac{1}{m_\pi}\mbox{\LARGE \{} 
                               \frac{5 g_{A}^2}{16} + \frac{g_{\pi 
                               N\Delta}^2}{\pi} \frac{m_\pi}{\Delta} \left[
                               1 + \frac{1}{9} \log \left( \frac{2\Delta}{
                               m_\pi} \right) \right] \nonumber \\
                         &   & \hspace{3cm} + O \left(\frac{m_{\pi}^3}{
                               \Delta^3}\right) \mbox{\LARGE \}} \\
\bar{\beta}^{\; \chi}_M  & = & \; + \;  \frac{e^2}{4\pi} \; \frac{1}{6\pi 
                               F_{\pi}^2} \; \frac{1}{m_\pi} \mbox{\LARGE \{} 
                               \frac{g_{A}^2}{32} + \frac{m_\pi}{\Delta} 
                               \left[ \frac{b_{1}^2}{3\pi} \frac{(4\pi 
                               F_\pi)^2}{M_{N}^2} + \frac{g_{\pi N\Delta}^2}{
                               9\pi} \log \left( \frac{2\Delta}{m_\pi} 
                               \right) \right] \nonumber \\
                         &   & \hspace{3cm} + O \left( \frac{m_{\pi}^3} 
                               {\Delta^3} \right) \mbox{\LARGE \}} 
\end{eqnarray}
We note that the long-range pion cloud, which scales as $1/m_\pi$,
provides the dominant singularity in the chiral limit, as expected.
Also, one can see that decoupling of the delta resonance in the chiral limit
is manifest.
Finally, we note that the leading order $p^4$-terms in the chiral expansion of
the $\Delta\pi$-loop contributions is already a good approximation to the full
order $\epsilon^3$ expressions. The contributions to $\bar{\alpha}_E$ and 
$\bar{\beta}_M$ are thereby changed by 15 and 7 \% respectively, indicating 
that the bulk of the effects due to $\Delta$(1232) are obtained at order 
$p^4$ in the chiral expansion ! Of course we emphasize that these 
considerations are only an {\em indication} of what might happen at the next 
order, 
future work will address these issues. We now move on to
discuss the case of spin-dependent quantities in Compton scattering.

\subsection{Spin-dependent Forward Compton Scattering}

In the presence of spin-dependence the Compton tensor for forward scattering 
of  real photons Eq.(\ref{eq:Amp}) from a spin 1/2 particle has to be expanded
by extra (spin-dependent) structures. Choosing $v_\mu =(1,0,0,0)$ and again 
working in Coulomb gauge $\epsilon \cdot v = \epsilon^\prime \cdot = 0$
we can write
\begin{eqnarray}
\epsilon^{\prime \; \mu} \; \Theta_{\mu\nu} \; \epsilon^{\nu} 
&=& e^{2} \mbox{\LARGE [ } 
\; - \; \hat{\epsilon}^\prime \cdot 
\hat{\epsilon} \; U(\omega) \; + \; \hat{\epsilon}^\prime \cdot 
\vec{k} \; \hat{\epsilon} \cdot \vec{k} \; V(\omega) \nonumber \\
& & + i \; \omega \; W^{(1)}(\omega) \; \vec{\sigma} \cdot \left( 
    \hat{\epsilon}^\prime \times \hat{\epsilon} \right)  
    + i \; \omega \; W^{(2)}(\omega) \; \hat{\epsilon} \cdot \vec{k} 
    \;\vec{\sigma} \cdot \left( \hat{\epsilon}^\prime \times \vec{k} 
    \right) \nonumber \\
& & + i \; \omega \; W^{(3)}(\omega) \; \hat{\epsilon}^\prime \cdot 
    \vec{k} \; \vec{\sigma} \cdot \left( \hat{\epsilon} \times \vec{k} 
    \right) \mbox{\LARGE ] } \; ,
\end{eqnarray}
where $\omega=v \cdot k$ denotes the energy of the forward scattering photon 
with four-momentum $k_\mu$. In the following we enforce the transversality
condition $\hat{\epsilon} \cdot \vec{k}=\hat{\epsilon}^\prime \cdot 
\vec{k}=0$ and therefore only work with the auxiliary 
function\footnote{From $W^{(2)}(\omega)$ and $W^{(3)}(\omega)$ one can obtain 
the third-order spin polarizabilities $\gamma_3 , \; \gamma_4$ of ref.
\cite{Ragusa}. However, we relegate this analysis to future work where we will
study the complete set of third-order spin polarizabilities of the nucleon.}
$W^{(1)}(\omega)$, which contains
the information about the spin 1/2 structure of the target nucleon. Having 
factored out an extra $\omega$,
we note that $W^{(1)}(\omega)$ is even under crossing (i.e. 
$W^{(1)}(\omega)= W^{(1)}(-\omega)$) and receives contributions
from four different sources, analogously to Eq.(\ref{eq:U4}):
\begin{equation}
W^{(1)}(\omega) = W^{(1)}_{N}(\omega)^{Born} + W^{(1)}_{N}(\omega)^{loops} +
                    W^{(1)}_{\Delta}(\omega)^{Born} + 
                    W^{(1)}_{\Delta}(\omega)^{loops} 
\end{equation}

First we calculate the nucleon Born contributions to $W^{(1)}(\omega)$. They 
arise at at ${\cal O}(\epsilon^3)$ from the four diagrams 2.1a-d in Figure 2. 
One finds
\begin{equation}
W^{(1)}_{N}(\omega)^{Born} = - \frac{1}{4 M_{N}^{2}} \; \left\{ \kappa_{p}^{2} 
                             \; \left( 1 + \tau_{3} \right) \; + \; 
                             \kappa_{n}^{2} \; \left( 1 - \tau_{3} \right) 
                             \right\} \label{eq:f2NBorn} \; ,
\end{equation}
where $\kappa_{p,(n)}$ corresponds to the anomalous magnetic moment for a
proton (neutron) target. As noted before, the contribution from the anomalous 
process $\pi^0 \rightarrow \gamma\gamma$ (diagram 2.1d in Figure 2) vanishes 
in the forward direction. However, the sum
of the other three diagrams satisfies the famous LET due to Gell-Mann, 
Goldberger and Low \cite{GMGL}, with the two-photon seagull diagram 2.1c
arising from $1/M^2$ corrections (see Table 1) playing a pivotal role.
Relating $W^{(1)}_N(\omega)^{Born}$ of Eq.(\ref{eq:f2NBorn}) to the function 
$f_2 (\omega)$ of Eq.(\ref{eq:f1f2}) via
$f_2(\omega)=e^2 /4\pi \; W^{(1)}(\omega)$, one finds in the limit
of zero photon energy
\begin{equation}
f_{2}(0)=-\; \frac{e^2\; \kappa^2}{8\pi \;M_N^2}
\end{equation}
Having satisfied the low energy constraint we conclude that
\begin{equation}
W^{(1)}_{N}(0)^{loop}=W^{(1)}_{\Delta}(0)^{Pole}=W^{(1)}_{\Delta}(0)^{loop}=0 \label{eq:cW} \; ,
\end{equation}
which constitutes a non-trivial check on our calculations.

The ${\cal O}(\epsilon^3)$ $N\pi$-loop contribution to $W^{(1)}(\omega)$ can 
be obtained from just six diagrams (3a-3f) of Figure 3. With details of the 
calculation given in Appendix B we find\footnote{Eq.(\ref{eq:WN})
agrees with the ${\cal O}(p^3)$ result of \cite{BKKM92}.} 
\begin{equation}
W^{(1)}_{N}(\omega)^{loops} = \frac{g_{A}^{2} }{2F_{\pi}^{2}}{\omega^2\over 
                        12\pi^2 m_\pi^2} + \; {\cal O}(\omega^4) \; ,
                        \label{eq:WN}
\end{equation}
so that the LET constraint $W^{(1)}_N(0)^{loop}=0$ is 
satisfied.

The ${\cal O}(\epsilon^3)$ delta Born contributions to the spin-flip function 
$W^{(1)}(\omega)$ are given by diagrams 2.2a and 2.2b of Figure 2, yielding
\begin{equation}
W^{(1)}_{\Delta}(\omega)^{Born}=-\; \frac{4 b_{1}^2}{9 M_{N}^2} \;
                                 \frac{\omega^2}{\Delta^2-\omega^2} \; .
\end{equation}
Again, the LET stricture Eq.(\ref{eq:cW}) holds explicitly.

Finally, the ${\cal O}(\epsilon^3)$ $\Delta\pi$-loop contributions to 
$W^{(1)}(\omega)$ arise from diagrams 4a-f in Figure 4. With details of 
the calculation given in Appendix C, we find 
\begin{eqnarray}
W^{(1)}_{\Delta}(\omega)^{loop}&=&\frac{g_{\pi N\Delta}^2 \; \omega^2}{54\pi^2
                                  \; F_{\pi}^2}\left[ \frac{3m_\pi^2\Delta}{
                                  (\Delta^2-m_\pi^2)^{5/2}} \log R - \frac{
                                  \Delta^2+2m_\pi^2}{(\Delta^2-m_\pi^2)^2}
                                  \right] \nonumber \\
& & +\; {\cal O}(\omega^4) \; ,
\end{eqnarray}
where $R$ has been defined in Eq.(\ref{eq:R}). The d-dependence of the 
pertinent amplitudes in Appendix C--induced by
the spin 3/2 propagator of Eq.(\ref{eq:propagator})--leads to a highly complex
cancelation pattern among the diagrams, ultimately yielding 
$W^{(1)}_\Delta(0)^{loop}=0$ as required by Eq.(\ref{eq:cW}).

In analogy to the spin-independent discussion in section 4.1 we introduce the
nucleon Born-term subtracted function
\begin{equation}
\hat{W}^{(1)}(\omega)=W^{(1)}(\omega) - W^{(1)}_N(\omega)^{Born}
\end{equation} 
and obtain the spin polarizability $\gamma$ of the nucleon
\begin{equation}
\gamma= \frac{e^2}{8\pi} \; \frac{\partial^2}{\partial \omega \;^2} 
\hat{W}^{(1)}(\omega)|_{\omega =0} \; ,
\end{equation}
which was defined via Eq.(\ref{eq:gamma}).

Summing the three contributions\footnote{Our findings agree with the 
$N\pi$- and $\Delta\pi$-loop results given in \cite{BKKM92}.} contained
in $\hat{W}^{(1)}(\omega)$ one finds
$\gamma$ to ${\cal O}(\epsilon^3)$:
\begin{eqnarray}
\gamma^{O(\epsilon^3)} & = & + \; \frac{e^2}{4\pi} \; \frac{g_{A}^2}{24 \pi^2 
                           F_{\pi}^{2}} \; \frac{1}{m_{\pi}^2} \nonumber \\
                     &   & - \; \frac{e^2}{4\pi} \; \frac{4 b_{1}^2}{9 M_{
                           N}^2} 
                           \; \frac{1}{\Delta^2} \nonumber \\ 
                     &   & - \; \frac{e^2}{4\pi} \; \frac{g_{\pi N\Delta}^2}{
                           54 \pi^2 F_{\pi}^2} \left[ \; \frac{\Delta^2
                           + 2 m_{\pi}^2}{(\Delta^2 - m_{\pi}^2)^2} 
                           - \; \frac{3 m_{\pi}^2 \Delta}{
                           (\Delta^2 - m_{\pi}^2)^{5/2}} \log R \; \right] 
                           \label{eq:g3} \\
                     & = & (4.5\mbox{ (N-loop)}-4.0\mbox{(delta-pole)}-0.4
                           \mbox{(delta-loop)})\times 10^{-4} \; {\rm fm}^4
                           \; . \nonumber 
\end{eqnarray}
We note that as in the case of $\bar{\alpha}_E$ and $\bar{\beta}_M$ 
there is no isospin dependence at ${\cal O}(\epsilon^3)$.
Also, $\Delta\pi$-loops are only playing a minor role
in the spin polarizability at this order. 
Finally, the large positive contribution
of the $N\pi$-loops is nearly completely canceled by the delta Born graphs.
In the case of the spin polarizability we cannot yet make a direct 
experimental comparison, as explained
in section 2.  However, 
comparing with the sum rule value given in Eq.(\ref{eq:dhg})
we see that the $\Delta(1232)$ 
contribution goes in the right direction but is not
large enough in (negative) magnitude in order to bring about
experimental agreement. We also note that at present we do not have a 
${\cal O}(p^4)$ calculation in HBChPT with which to compare.
However, the one-loop relativistic 
baryon results of Eq.(\ref{eq:gr}) already give an indication that even if 
one extends the $O(p^3)$ HBChPT result of Eq.(\ref{eq:gg3}) to the next 
order, one will not be able to describe $\gamma$ without keeping 
$\Delta$(1232) as an
explicit degree of freedom in the theory. 
Studying the chiral limit of the spin polarizability supports this 
viewpoint. We find
\begin{eqnarray}
\gamma^\chi &=& \frac{e^2}{4\pi} \; \frac{1}{216\pi^2 \; F_{\pi}^2} \;
\frac{1}{m_{\pi}^2} \left[ 9 g_{A}^2 - 6 b_{1}^2 \; \frac{(4\pi F_\pi)^2}{
M_{N}^2} \; \frac{m_{\pi}^2}{\Delta^2} - 4 g_{\pi N\Delta}^2 \; 
\frac{m_{\pi}^2}{\Delta^2} + \; {\cal O}\left(\frac{m_{\pi}^4}{\Delta^4}
\right) \; \right] \nonumber \\
& & \label{eq:gch}
\end{eqnarray}
One observes that the $1/m_{\pi}^2$ singularity due to the pion cloud of the
nucleon remains leading in the chiral
limit and that delta decoupling holds. We
also note that the $\Delta\pi$-loop contributions increase in magnitude
to $- 0.6 \; 10^{-4} \; {\rm fm}^4$ but remain small.
Most importantly, Eq.(\ref{eq:gch}) shows us that all delta effects only
start contributing at ${\cal O}(p^5)$ (!) in standard SU(2) HBChPT. The
spin dependent polarizability of the nucleon is therefore a prime
example of how a theory with explicit delta degrees of freedom can 
dramatically improve the convergence of the perturbation series. 
In light of the cancelations in 
Eq.(\ref{eq:g3}) we are encouraged that at ${\cal O}(\epsilon^4)$ 
one may be able to achieve  a good understanding of the spin 
polarizability with the help of the small scale expansion\footnote{
There certainly have been many attempts in the literature to call the delta
to the rescue in order to get reasonable numbers for the sum rules of 
Eq.(\ref{eq:dhg}). We emphasize again that our results in Eq.(\ref{eq:g3})
follow from the {\em systematic small scale expansion in HBChPT}, as laid
out in section 3.}.

\subsection{Critical Discussion of ${\cal O}(\epsilon^3)$ Results}

Before concluding it is useful to discuss some of the arguments of which
we are aware why an ${\cal O}(\epsilon^3)$ calculation can 
perform poorly on the phenomenological side and on the other hand
show all the correct field-theoretic constraints of LETs,
heavy mass decoupling, correct chiral limit etc.

i) The heavy baryon propagator for nucleons of Eq.(\ref{eq:propagator})
receives ``propagator corrections'' via 1/M vertices in
matrix $\tilde{{\cal A}}_{N}^{(2)}$ of Eq.(\ref{eq:Atilde}), which in 
one-loop diagrams only 
begin to show up at ${\cal O}(\epsilon^4)$. From a relativistic viewpoint
these $1/M$ corrections correspond to effects pertaining
to the lower components of a Dirac spinor.

ii) For the heavy baryon propagator of a spin 3/2 particle there are
analogous $1/M$ induced propagator corrections in matrix 
$\tilde{{\cal A}}_{\Delta}^{(2)}$ of Eq.(\ref{eq:Atilde}). In this case 
they have the additional
effect of correcting the pure spin 3/2 projector of Eq.(\ref{eq:propagator})
by bringing in some information about the off-shell spin 1/2 components
of a relativistic spin 3/2 particle. Phenomenologically it is well
known that the spin 1/2 components of the relativistic spin 3/2
propagator can play an important effect in some observables \cite{benm}.
Again, at the one-loop level these effects only start showing up
at ${\cal O}(\epsilon^4)$.

iii) In ${\cal O}(\epsilon^3)$ calculations the incoming (and outgoing)
soft momentum of the baryon is usually a higher order effect.
Only at the next order does this momentum fully contribute and the size 
of these ``recoil corrections'' can be quite large for some observables.

iv) Isospin is usually not broken in calculations at the 
${\cal O}(\epsilon^3)$ level which can be a problem in some cases.

v) Often one is calculating observables which are dominated by loop
effects. Therefore one is only calculating the leading order of these
observables, although technically one might have to work to third or fourth
order in HBChPT. As experience from many calculations in the meson sector
of ChPT teaches us, one should always consider the leading and the first
(at least) non-leading order in an observable before one can try to judge
the quality of a ChPT result.

This list is certainly not complete. The concerns raised are valid beyond our 
particular case 
of the ${\cal O}(\epsilon^3)$ calculation of the polarizabilities of 
the nucleon. In fact, most of these points also apply to SU(2) HBChPT without
explicit delta degrees of freedom and ``leading log'' calculations in
SU(3) HBChPT. For our particular case of interest---the polarizabilities
of the nucleon---all of these points strongly suggest moving onto the
${\cal O}(\epsilon^4)$ calculation.

\section{Conclusions}

The Compton scattering process offers the opportunity to probe nucleon
structure in a relatively clean fashion, via measurement of various
``polarizabilities'' which probe the nucleon response to quasi-static
excitations.   In particular recent years have seen the measurement of
electric and magnetic polarizabilities for both neutron and proton and
spin polarizability measurements should be available by the millennium.
On the theoretical side, ${\cal O}(p^3)$ predictions for
$\bar{\alpha}_E$ and $\bar{\beta}_M$ within heavy baryon chiral
perturbation theory give a surprisingly good picture of the
experimental situation.  However, corresponding predictions for the
spin polarizability are not in good agreement with the values given
from DHG sum rule arguments. When extended to ${\cal O}(p^4)$ reasonable
agreement is obtained for $\bar{\alpha}_E$ and $\bar{\beta}_M$, but at the 
cost of
considerable uncertainty associated with effects such as the
$\Delta(1232)$, which is included only as a very heavy particle
contributing to various counterterms.  In this paper we have removed this
obstacle by treating the $\Delta(1232)$ as a specific degree of
freedom within the heavy baryon method.  Our calculation was performed
to ${\cal O}(\epsilon^3)$ where $\epsilon$ is taken as a soft momentum,
as $m_\pi$ or as the nucleon-delta mass difference.  We find, perhaps not
surprisingly, that inclusion of delta effects makes large changes to
all three polarizabilities. 

The good agreement with experiment at 
${\cal O}(p^3)$ for $\bar{\alpha}_E$ and $\bar{\beta}_M$ is destroyed at
${\cal O}(\epsilon^3)$, whereas the spin polarizability $\gamma$ improves
dramatically when compared with currently available sum rule information.
Furthermore, we have discussed consequences of our HBChPT results in the 
light of  
existing ${\cal O}(p^4)$ and relativistic one-loop calculations.
Indeed, we regard our calculation as
merely preliminary and look forward to extending this work to higher order.

\newpage
\appendix

\section{Loop Functions}

We express the invariant amplitudes of Feynman diagrams containing pion-loops
around a nucleon in terms of J-functions, defined via
\begin{eqnarray}
{1\over i}\int{d^d\ell\over (2\pi)^d}
{\{1,\ell_\mu\ell_\nu,\ell_\mu\ell_\nu\ell_\alpha\ell_\beta\}\over 
(v\cdot\ell-\omega-i\epsilon)(m_\pi^2-\ell^2-i\epsilon)}&=&
\{J_0(\omega,m_\pi^2),\\
& & g_{\mu\nu}J_2(\omega,m_\pi^2)+v_\mu v_\nu J_3(\omega,m_\pi^2),\nonumber\\
& & (g_{\mu\nu} g_{\alpha\beta}+{\rm perm.})J_6(\omega,m_\pi^2)+ ... \nonumber
\}
\end{eqnarray}
Employing the following identities
\begin{eqnarray}
J_2 \left( \omega, m_{\pi}^2 \right) & = & \frac{1}{d-1} \left[ \left(
    m_{\pi}^2 - \omega^2 \right) J_0 \left( \omega, m_{\pi}^2 \right)
    - \omega \; \Delta_{\pi} \right] \label{J_2}\\
J_6 \left( \omega, m_{\pi}^2 \right) & = & \frac{1}{d+1} \left[ \left(
    m_{\pi}^2 - \omega^2 \right) J_2 \left( \omega , m_{\pi}^2 \right)
    - \frac{m_{\pi}^2 \omega}{d} \; \Delta_\pi \right] \label{J_6} \, ,
\end{eqnarray}
one concludes that all loop-integrals can be expressed via
the basis-function $J_0$. For the $N\pi$ loop-integrals we use \cite{BKM95}
\begin{equation}
J_{0} \left( \omega, m_{\pi}^2 \right) =  - 4 L \omega +
\frac{\omega}{8 \pi^2} \left( 1 - 2 {\mathrm{ln}} \frac{m_{\pi}}{\mu}
\right) - \frac{1}{4 \pi^2} \sqrt{m_{\pi}^2 - \omega^2}
{\mathrm{arccos}} \frac{-\omega}{m_{\pi}} \, , \label{J_0} 
\end{equation}
whereas for the $\Delta\pi$ loop-integrals we employ the analytically 
continued function
\begin{eqnarray}
J_{0} \left( \omega, m_{\pi}^2 \right) &=&  - 4 L \omega +
\frac{\omega}{8 \pi^2} \left( 1 - 2 {\mathrm{ln}} \frac{m_{\pi}}{\mu}
\right) + \frac{1}{4 \pi^2} \sqrt{\omega^2 - m_{\pi}^2} \nonumber \\
& & \hspace{3cm} \times \log \left[ - \; \frac{\omega}{m_\pi}+\sqrt{\frac{
\omega^2}{m_{\pi}^2}-1} \right] \; . \label{J_0a}
\end{eqnarray}
In Eqs.\ (\ref{J_2}), (\ref{J_6}), (\ref{J_0}) and (\ref{J_0a}) we have used 
the same conventions as \cite{BKM95},
\begin{eqnarray}
\Delta_{\pi} & = & 2 m_{\pi}^2 \left( L + \frac{1}{16 \pi^2}
  {\mathrm{ln}} \frac{m_{\pi}}{\mu} \right) + {{O}} \left( d - 4
\right) \, , \nonumber\\
L & = & \frac{\mu^{d-4}}{16 \pi^2} \left[ \frac{1}{d-4} + \frac{1}{2}
  \left( \gamma_E - 1 - {\mathrm{ln}} 4 \pi \right) \right] \, ,
\end{eqnarray}
where we introduced the Euler-Mascharoni constant, $\gamma_E =
0.557215$, and the scale $\mu$ in the dimensional regularization
scheme we use for the evaluation of the integrals.

Finally, with $J_{i}'$ and
$J_{i}''$ we define the first and second partial derivative with respect to
$m_{\pi}^2$,
\begin{eqnarray}
J_i'\left(\omega,m_{\pi}^2\right) & = & \frac{\partial}{\partial
\left( m_{\pi}^2 \right)} J_i \left( \omega,m_{\pi}^2 \right) \, , \\
J_i''\left(\omega,m_{\pi}^2\right) & = & \frac{\partial^2}{\partial
\left( m_{\pi}^2 \right)^2} J_i \left( \omega,m_{\pi}^2 \right) \, .
\end{eqnarray}

\newpage
\section{$N\pi$ Loops in Forward Compton Scattering}

Using the J-function formalism defined in Appendix A and the lagrangians of
Eq.(\ref{eq:llead}) one can get exact 
solutions for the nine $N\pi$-loop diagrams of Figure 3. With $\epsilon_\mu$
($\epsilon_{\mu}^{\; \prime}$) we denote the polarization four-vector of
the incoming (outgoing) photon with constant four-momentum $k_\mu$ and 
energy $\omega$. We find 
\begin{eqnarray}
Amp_{1+2}^{N\pi}  &=& C \; \bar{u}(r) \mbox{\LARGE \{ } 
                      - \epsilon \cdot \epsilon^\prime 
                      \left[ J_0 ( \omega , m_{\pi}^2 ) + 
                       J_0 ( - \omega , m_{\pi}^2 ) \right] \\
                  & & + 2 \; [ S_\mu , S_\nu ] \; \epsilon^{\prime \; \mu}
                      \epsilon^\nu \left[ J_0 ( \omega , m_{\pi}^2 ) -
                       J_0 ( - \omega , m_{\pi}^2 ) \right]
                      \mbox{\LARGE \} } u(r) \nonumber \\
Amp_{3..6}^{N\pi} &=& C \; \bar{u}(r) \mbox{\LARGE \{ } 
                      +4 \; \epsilon \cdot \epsilon^\prime \int_{o}^{1} dx  
                      \left[ J_{2}^\prime ( \omega x , m_{\pi}^2 ) + 
                       J_{2}^\prime ( - \omega x , m_{\pi}^2 ) \right] \\
                  & & -8 \; [ S_\mu , S_\nu ] \; \epsilon^{\prime \; \mu}
                      \epsilon^\nu \int_{o}^{1} dx  
                      \left[ J_{2}^\prime ( \omega x , m_{\pi}^2 ) -
                       J_{2}^\prime ( - \omega x , m_{\pi}^2 ) \right]
                      \nonumber \\
                  & & -2 \; \epsilon \cdot k \; \epsilon^\prime \cdot k 
                      \int_{0}^{1} dx \; x(1-2x) 
                      \left[ J_{0}^\prime ( \omega x , m_{\pi}^2 ) +
                       J_{0}^\prime ( - \omega x , m_{\pi}^2 ) \right]
                      \nonumber \\
                  & & +2 \;[ S_\mu , S_\nu ] \; \left( \epsilon^\prime \cdot k
                      \; k^\mu \epsilon^\nu + \epsilon \cdot k \; 
                      \epsilon^{\prime \; \mu} k^\nu \right) 
                      \int_{0}^{1} dx \; x(1-2x) \nonumber \\
                  & & \hspace{3cm} \times 
                      \left[ J_{0}^\prime ( \omega x , m_{\pi}^2 ) -
                       J_{0}^\prime ( - \omega x , m_{\pi}^2 ) \right] 
                      \mbox{\LARGE \} } u(r) \nonumber \\
Amp_{7+8}^{N\pi}  &=& C \; \bar{u}(r) \mbox{\LARGE \{ }
                      -4 \; \epsilon \cdot \epsilon^\prime \; (d+1) 
                      \int_{o}^{1} dx \; (1-x)  
                      \nonumber \\
                  & & \hspace{3cm} \times 
                      \left[ J_{6}^{\prime\prime} ( \omega x , 
                      m_{\pi}^2 ) +  J_{6}^{\prime\prime} ( - \omega x , 
                      m_{\pi}^2 ) \right] \\
                  & & +4 \; \epsilon \cdot \epsilon^\prime \; \omega^2 
                      \int_{o}^{1} dx \;(1-x) x^2 
                      \left[ J_{2}^{\prime\prime} ( \omega x , 
                      m_{\pi}^2 ) +  J_{2}^{\prime\prime} ( - \omega x , 
                      m_{\pi}^2 ) \right] \nonumber \\
                  & & - \; \epsilon \cdot k \; \epsilon^\prime \cdot k 
                      \int_{0}^{1} dx \; (1-x)
                      \left[ 8x (2x-1) + (2x-1)^2 (d-1) \right] \nonumber \\  
                  & & \hspace{3cm} \times 
                      \left[ J_{2}^{\prime\prime} ( \omega x , 
                      m_{\pi}^2 ) +  J_{2}^{\prime\prime} ( - \omega x , 
                      m_{\pi}^2 ) \right] \nonumber \\
                  & & + \; \epsilon \cdot k \; \epsilon^\prime \cdot k \; 
                      \omega^2 \int_{0}^{1} dx \;
                      (1-x) \; x^2 \; (2x-1)^2 \nonumber \\
                  & & \hspace{3cm} \times 
                      \left[ J_{0}^{\prime\prime} ( \omega x , 
                      m_{\pi}^2 ) +  J_{0}^{\prime\prime} ( - \omega x , 
                      m_{\pi}^2 ) \right]
                      \mbox{\LARGE \} } u(r) \nonumber \\
Amp_{9}^{N\pi}    &=& C \; \bar{u}(r) \mbox{\LARGE \{ } 
                      +2 \; \epsilon \cdot \epsilon^\prime \; (d-1) \; 
                      J_{2}^\prime ( 0 , m_{\pi}^2 )  
                      \mbox{\LARGE \} } u(r) \; ,
\end{eqnarray}
\newpage
with the common factor
\begin{eqnarray}
C= i \; \frac{e^2 g_{A}^2}{2 F_{\pi}^2} \; , \nonumber
\end{eqnarray} 
In order to evaluate the polarizabilities, we expand the nine amplitudes into
a power series in $\omega$ and only keep the terms of interest:
\begin{eqnarray}
Amp_{1+2}^{N\pi}  &=& C \; \bar{u}(r) \mbox{\LARGE \{ } 
                      + \; \epsilon \cdot \epsilon^\prime \left[ \frac{
                      m_\pi}{4\pi} - \omega^2
                      \; \frac{1}{8\pi m_\pi} + {\cal O}(\omega^4)  \right] 
                      \label{eq:a12} \\
                  & & + \omega \; [ S_\mu , S_\nu ] \; \epsilon^{\prime \; \mu}
                      \epsilon^\nu \mbox{\LARGE \bf{ [ }}- \; \frac{1}{2\pi^2} 
                      \left( 32\pi^2 L + 1 + 2 \log \frac{m_\pi}{\lambda} 
                      \right)  \nonumber \\
                  & & \hspace{3.7cm} + \; \omega^2 \; \frac{1}{3\pi^2 
                      m_{\pi}^2} + {\cal O}(\omega^4) \mbox{\LARGE \bf{ ] }}
                      \mbox{\LARGE \} } u(r) \nonumber \\
Amp_{3..6}^{N\pi} &=& C \; \bar{u}(r) \mbox{\LARGE \{ } 
                      + \; \epsilon \cdot \epsilon^\prime \left[ - \; \frac{
                      m_\pi}{2\pi} +
                      \omega^2 \; \frac{1}{12 \pi m_\pi} + {\cal O}(\omega^4) 
                      \right] \\
                  & & + \omega \; [ S_\mu , S_\nu ] \; \epsilon^{\prime \; \mu}
                      \epsilon^\nu \mbox{\LARGE \bf{ [ }} \frac{1}{2\pi^2} 
                      \left( \frac{96\pi^2}{(d-1)} L + \frac{5}{3} + 2 \log 
                      \frac{m_\pi}{\lambda} \right) \nonumber \\
                  & & \hspace{3.7cm} - \; \omega^2 \; \frac{1}{6\pi^2 
                      m_{\pi}^2} + {\cal O}(\omega^4) \mbox{\LARGE \bf{ ] }}
                      \nonumber \\
                  & & + \; \epsilon \cdot k \; \epsilon^\prime \cdot k  
                      \left[ - \; \frac{1}{24\pi 
                      m_\pi} + {\cal O}(\omega^4) \right] \nonumber \\
                  & & + \; ... 
                      \mbox{\LARGE \} } u(r) \nonumber \\
Amp_{7+8}^{N\pi}  &=& C \; \bar{u}(r) \mbox{\LARGE \{ }
                      + \; \epsilon \cdot \epsilon^\prime  \left[ \frac{5 
                      m_\pi}{8\pi}- \omega^2
                      \; \frac{5}{96\pi m_\pi} + {\cal O}(\omega^4) \right] \\
                  & & + \; \epsilon \cdot \epsilon^\prime  \left[ - \omega^2 
                      \; \frac{2}{96\pi 
                      m_\pi} + {\cal O}(\omega^4) \right] \nonumber \\
                  & & + \; \epsilon \cdot k \; \epsilon^\prime \cdot k  
                      \left[ \frac{1}{32\pi m_\pi}
                      + {\cal O}(\omega^2) \right] \nonumber \\  
                  & & + \; ... \mbox{\LARGE \} } u(r) \nonumber \\
Amp_{9}^{N\pi}    &=& C \; \bar{u}(r) \mbox{\LARGE \{ } 
                      + \; \epsilon \cdot \epsilon^\prime \left[ - \frac{3 
                      m_\pi}{8 \pi} \right]  
                      \mbox{\LARGE \} } u(r) \label{eq:a9} 
\end{eqnarray}
In section 4 we construct the auxiliary functions $U(\omega), \; V(\omega)$
and $W(\omega)$ from Eqs.(\ref{eq:a12}-\ref{eq:a9}).
\newpage

\section{$\Delta\pi$ Loops in Forward Compton Scattering}

Using the same conventions as in Appendix B with a new overall factor
\begin{eqnarray}
D=i \; \frac{4 e^2 g_{\pi N\Delta}^2}{3 F_{\pi}^2} \; , \nonumber
\end{eqnarray}
we find the invariant amplitudes for the nine diagrams of Figure 4:
\begin{eqnarray}
Amp^{\Delta\pi}_{1+2} &=& D \; \bar{u}(r) \mbox{ \LARGE \{ } 
                       - \;\epsilon \cdot \epsilon^\prime \; \frac{d-2}{d-1} 
                       \left[ J_0 ( \omega - 
                       \Delta , m_{\pi}^2 ) + J_0 ( - \omega - \Delta , 
                       m_{\pi}^2 ) \right] \\
                   & & - \; [ S_\mu \; , \; S_\nu ] \; 
                       \epsilon^{\prime \mu}
                       \epsilon^\nu \; \frac{2}{d-1} \left[ J_0 ( \omega - 
                       \Delta , m_{\pi}^2 ) - J_0 ( - \omega - \Delta , 
                       m_{\pi}^2 ) \right] \mbox{ \LARGE \} } u(r) \nonumber \\
Amp^{\Delta\pi}_{3..6}&=& D \; \bar{u}(r) \mbox{ \LARGE \{ } 
                       4 \;\epsilon \cdot \epsilon^\prime \; \frac{d-2}{d-1} 
                       \int_{0}^{1} dx
                       \left[ J_{2}^\prime ( \omega x - \Delta , m_{\pi}^2 ) +
                       J_{2}^\prime ( - \omega x - \Delta , m_{\pi}^2 ) 
                       \right] \\
                   & & + \; 4 \; [ S_\mu \; , \; S_\nu ] \; 
                       \epsilon^{\prime \mu} \epsilon^\nu \; \frac{2}{d-1} 
                       \int_{0}^{1} dx \left[ J_{2}^\prime ( \omega x - \Delta
                       , m_{\pi}^2 ) - J_{2}^\prime ( - \omega x - \Delta , 
                       m_{\pi}^2 ) \right] \nonumber \\
                   & & - \; 2 \; \epsilon \cdot k \; \epsilon^\prime \cdot k 
                       \; \frac{d-2}{d-1} \int_{0}^{1} dx \; x(1-2x) 
                       \left[ J_{0}^\prime ( \omega x -
                       \Delta , m_{\pi}^2 ) + J_{0}^\prime ( - \omega x - 
                       \Delta , m_{\pi}^2 ) \right] \nonumber \\
                   & & - \; [ S_\mu \; , \; S_\nu ] 
                       \left( \epsilon^\prime \cdot k \; k^\mu \epsilon^\nu 
                       + \epsilon \cdot k \; \epsilon^{\prime \mu} k^\nu
                       \right) \; \frac{2}{d-1} \int_{0}^{1} dx \; x(1-2x)
                       \nonumber \\
                   & & \hspace{3cm} \times \; \left[ J_{0}^\prime ( \omega x
                       - \Delta , m_{\pi}^2 ) - J_{0}^\prime ( - \omega x - 
                       \Delta , m_{\pi}^2 ) \right]  \mbox{ \LARGE \} } u(r) 
                       \nonumber \\
Amp^{\Delta\pi}_{7+8} &=& D \; \bar{u}(r) \mbox{ \LARGE \{ } 
                       - \; \epsilon \cdot \epsilon^\prime \; 4 \; \frac{
                       (d-2)(d+1)}{(d-1)} \;
                       \int_{0}^{1} dx \; (1-x) \\
                   & & \hspace{3cm} \times \; \left[ J_{6}^{\prime\prime} 
                       ( \omega x - \Delta , m_{\pi}^2 ) + J_{6}^{\prime
                       \prime}( - \omega x - \Delta , m_{\pi}^2 ) \right]
                       \nonumber \\
                   & & + \; \epsilon \cdot \epsilon^\prime \; \omega^2 \; 4 
                       \; \frac{d-2}{d-1} \;
                       \int_{0}^{1} dx \; (1-x) \; x^2 \left[ J_{2}^{\prime
                       \prime} ( \omega x - \Delta , m_{\pi}^2 ) + J_{2}^{
                       \prime\prime}( - \omega x - \Delta , m_{\pi}^2 ) \right]
                       \nonumber \\
                   & & - \; \epsilon \cdot k \; \epsilon^\prime \cdot k \; 
                       \frac{d-2}{d-1} \;
                       \int_{0}^{1} dx \; (1-x) \; (2x-1) \; [8x+(2x-1)(d-1)]
                       \nonumber \\
                   & & \hspace{3cm} \times \; \left[ J_{2}^{\prime\prime} 
                       ( \omega x - \Delta , m_{\pi}^2 ) + J_{2}^{\prime
                       \prime}( - \omega x - \Delta , m_{\pi}^2 ) \right]
                       \nonumber \\
                   & & + \; \epsilon \cdot k \; \epsilon^\prime \cdot k \; 
                       \omega^2 \; \frac{d-2}
                       {d-1} \; \int_{0}^{1} dx \; (1-x) \; x^2 \; (2x-1)^2 
                       \nonumber \\
                   & & \hspace{3cm} \times \; \left[ J_{0}^{\prime\prime} 
                       ( \omega x - \Delta , m_{\pi}^2 ) + J_{0}^{\prime
                       \prime}( - \omega x - \Delta , m_{\pi}^2 ) \right]
                       \mbox{ \LARGE \} } u(r) \nonumber \\
Amp^{\Delta\pi}_9  &=& D \; \bar{u}(r) \; \epsilon \cdot \epsilon^\prime \; 
                       2 \; (d-2) \;
                       J_{2}^\prime ( -\Delta , m_{\pi}^2 ) \; u(r) 
\end{eqnarray}
\newpage
In analogy to Appendix B we expand the amplitudes into a power-series in the
photon energy $\omega$. With the definition
\begin{eqnarray}
R=\frac{\Delta}{m_\pi}+\sqrt{\frac{\Delta^2}{m_{\pi}^2}-1} \nonumber 
\end{eqnarray}
we find: 
\begin{eqnarray}
Amp^{\Delta\pi}_{1+2} &=& D \; \bar{u}(r) \mbox{ \LARGE \{ } 
                       - \epsilon \cdot \epsilon^\prime \; \frac{2}{3} 
                       \mbox{ \LARGE [ } \frac{3}{2} \; \frac{d-2}{d-1} \;2 \; 
                       J_0 ( - \Delta , m_{\pi}^2 ) \\
                   & & \hspace{3.5cm} + \frac{\omega^2}{4 \pi^2}
                       \left( \frac{\Delta}{\Delta^2 - m_{\pi}^2} 
                       - \; \frac{m_{\pi}^2}{(\Delta^2 - m_{
                       \pi}^2)^{3/2}} \; \log R \right) + {\cal O}(\omega^4) 
                       \mbox{ \LARGE ] } \nonumber \\
                   & & + \; [ S_\mu \; , \; S_\nu ] \;
                       \epsilon^{\prime \mu} \epsilon^\nu \;  
                       \frac{\omega}{6 \pi^2} \mbox{ \LARGE [ } \frac{3}{d-1} 
                       \; 32\pi^2 L + 1 + 2 
                       \; \log ( \frac{m_\pi}{\lambda} ) \nonumber \\
                   & & \hspace{3.5cm} + \frac{2 \Delta}{\sqrt{\Delta^2 -
                       m_{\pi}^2}} \log R + \; \omega^2 \mbox{ \LARGE ( } 
                       \frac{\Delta m_{\pi}^2}{(\Delta^2 - m_{\pi}^2)^{5/2}}
                       \; \log R \nonumber \\
                   & & \hspace{3.5cm} - \; \frac{\Delta^2 + 2 m_{\pi}^2}{3 \; (
                       \Delta^2 -m_{\pi}^2)^2} \mbox{ \LARGE ) } + {\cal O}
                       (\omega^4) \mbox{ \LARGE ] } \mbox{ \LARGE \} } u(r)
                       \nonumber \\
Amp^{\Delta\pi}_{3..6}&=& D \; \bar{u}(r) \mbox{ \LARGE \{ }  
                       \epsilon \cdot \epsilon^\prime \; \frac{1}{9 \pi^2} 
                       \mbox{ \LARGE [ }
                       \frac{9}{2} \; \frac{d-2}{(d-1)^2} \; 96 \Delta L \pi^2
                       + 6 \Delta \log (\frac{m_\pi}{\lambda})   \\
                   & & \hspace{3.5cm} - \Delta + 6 \sqrt{\Delta^2 - m_{\pi}^2}
                       \; \log R + \omega^2 \mbox{ \LARGE ( }
                       \frac{\Delta}{\Delta^2 -m_{\pi}^2} \nonumber \\
                   & & \hspace{3.5cm} - \frac{m_{\pi}^2}
                       {(\Delta^2 -m_{\pi}^2)^{3/2}} 
                       \log R \mbox{ \LARGE ) } + {\cal O}(\omega^4)
                       \mbox{ \LARGE ] } \nonumber \\
                   & & + \; [ S_\mu \; , \; S_\nu ] \; 
                       \epsilon^{\prime \mu} \epsilon^\nu \;  
                       \frac{\omega}{36 \pi^2} \mbox{ \LARGE [ }  
                       9 \; \frac{-2}{(d-1)^2} \; 96 \pi^2 L - 12
                       \log ( \frac{m_\pi}{\lambda} ) - 10 \nonumber \\
                   & & \hspace{3.5cm} - \frac{12 \Delta}{\sqrt{\Delta^2 -
                       m_{\pi}^2}} \; \log R
                       + \omega^2 \mbox{ \LARGE ( } \frac{
                       \Delta^2 + 2 m_{\pi}^2}{(\Delta^2 - m_{\pi}^2)^2}
                       \nonumber \\
                   & & \hspace{3.5cm} - \frac{3 \Delta m_{\pi}^2}{(\Delta^2 -
                       m_{\pi}^2)^{5/2}} \; \log R \mbox{ \LARGE ) }
                       + {\cal O}(\omega^4) \mbox{ \LARGE ] } \nonumber \\
                   & & - \; \epsilon \cdot k \; \epsilon^\prime \cdot k \; 
                       \frac{1}{18 \pi^2} \mbox{ \LARGE [ } \frac{1}{
                       \sqrt{\Delta^2 -m_{\pi}^2}} \; \log R + 
                       {\cal O}(\omega^2) \mbox{ \LARGE ] } \nonumber \\
                   & & + \; ... \mbox{ \LARGE \} } u(r) \nonumber \\
Amp^{\Delta\pi}_{7+8}&=& D \; \bar{u}(r) \mbox{ \LARGE \{ } 
                       - \; \epsilon \cdot \epsilon^\prime \; \frac{1}{72 
                       \pi^2} \; 
                       \mbox{ \LARGE [ } \frac{18 (d-2)}{d \; (d-1)^2} \;
                       64 (4d-1) \Delta L \pi^2  \\
                   & & \hspace{4cm} + 60 \Delta
                       \log (\frac{m_\pi}{\lambda} ) + 60 \sqrt{\Delta^2
                       - m_{\pi}^2} \; \log R + \Delta \nonumber \\
                   & & \hspace{4cm} + \omega^2  
                       \left( \frac{5 \Delta}{\Delta^2 -m_{\pi}^2}
                       - \frac{5 m_{\pi}^2}{(\Delta^2 -m_{\pi}^2)^{3/2}} \;
                       \log R \right) + {\cal O}(\omega^4)
                       \mbox{ \LARGE ] } \nonumber \\
                   & & - \; \epsilon \cdot \epsilon^\prime \; \omega^2 \; 
                       \frac{1}{36 \pi^2} \mbox{ \LARGE [ }
                       \frac{1}{\sqrt{\Delta^2 -m_{\pi}^2}} \; \log R + 
                       {\cal O}(\omega^2) \mbox{ \LARGE ] }
                       \nonumber \\
                   & & + \; \epsilon \cdot k \; \epsilon^\prime \cdot k \; 
                       \frac{1}{24 \pi^2} \mbox{ \LARGE [ }
                       \frac{1}{\sqrt{\Delta^2 -m_{\pi}^2}} \; \log 
                       R + {\cal O}(\omega^2) \mbox{ \LARGE ] }
                       \nonumber \\
                   & & + \; ... \mbox{ \LARGE \} } u(r) \nonumber \\
Amp^{\Delta\pi}_9  &=& D \; \bar{u}(r) \; \epsilon \cdot \epsilon^\prime 
                       \; \frac{1}{12 \pi^2}
                       \mbox{ \LARGE \{ } \frac{3}{2} \; \frac{d-2}{d-1} \;
                       96 \Delta L \pi^2 + 6 \Delta \log (\frac{m_\pi}{\lambda}
                       ) - \Delta \\
                   & & \hspace{4cm} + 6 \sqrt{\Delta^2 - m_{\pi}^2} \;
                       \log R \mbox{ \LARGE \} } u(r) \nonumber
\end{eqnarray}

\newpage

\begin{figure}
{\bf FIGURE CAPTIONS:}

Wiggly lines to the right (left) denote incoming (outgoing) photons, dotted 
lines denote pions and solid lines represent nucleons.
\caption{\label{fig:b2} O($\epsilon^2$) Born graph in forward Compton
scattering.}
\caption{\label{fig:b3} O($\epsilon^3$) Born contributions in forward Compton
scattering.}
\caption{\label{fig:Npi} O($\epsilon^3$) $N\pi$ Loop diagrams in forward 
Compton scattering.}
\caption{\label{fig:Dpi} O($\epsilon^3$) $\Delta\pi$ Loop diagrams  in forward
 Compton scattering.}
\end{figure}
\vfill

\end{document}